\documentclass[12pt]{article}
\usepackage{amssymb,amsmath,epsfig}

\renewcommand{\theequation}{\arabic{section}.\arabic{equation}}

\begin{document}

\title{\bf Dispersion Modes of Hot Plasma for Schwarzschild de-Sitter Horizon in a Veselago Medium}
\author{M. Sharif \thanks{msharif.math@pu.edu.pk} and Ifra Noureen
\thanks{ifra.noureen@gmail.com}\\
Department of Mathematics, University of the Punjab,\\
Quaid-e-Azam Campus, Lahore-54590, Pakistan.}
\date{}
\maketitle
\begin{abstract}
We analyze the dispersion modes of hot plasma around Schwarzschild
de-Sitter event horizon in the presence of Veselago medium. For this
purpose, we apply ADM $3+1$ formalism to develop GRMHD equations for
the Schwarzschild de-Sitter spacetime. Implementation of linear
perturbation yields perturbed GRMHD equations that are used for the
Fourier analysis of rotating (non-magnetized and magnetized) plasma.
Wave analysis is described by the graphical representation of the
wave vector, refractive index, change in refractive index, phase and
group velocities. The outcome of this work supports the validity of
Veselago medium.
\end{abstract}

\section{Introduction}

Plasma is a distinct state of matter with a collection of free
electric charge carriers which behave collectively and respond
strongly to electromagnetic fields. Debye shielding effect is the
common example of collective effects of the plasma particles. The
number density of positively charged particles remains almost the
same as that of negatively charged particles to preserve the state
of quasi-neutrality. Plasma is formed by the recombination of free
electrons and ions, so it must have thermal energies to overcome the
coupling of charged particles and can interact with electromagnetic
fields. The nature of particle interactions distinguish between
weakly and strongly ionized plasma. In weakly ionized plasma,
charge-neutral interactions dominate the multiple Coulomb
interactions, otherwise it is strongly ionized \cite{1}.
It is mentioned here that the most general plasma is hot and highly
ionized \cite{2}.

Black holes are amongst the mysterious compact objects, depicting
region of space with strong gravitational field that nothing not
even light can escape from it. According to the predictions of
relativists, black holes are formed when matter is compressed to a
point. These are commonly created from deaths of massive stars or by
collapse of a supergiant star. Black holes contain the energy flux
which generates a relatively large magnetic field \cite{3}.
The theory of magnetohydrodynamics (MHD) is developed for the
study of plasma flow having both positive and negative charge
carriers. This describes the study of plasma flow in the presence of
magnetic field. The MHD theory along with the effects of gravity is
said to be general relativistic magnetohydrodynamics (GRMHD). The
strong gravity of black hole perturbs the magnetospheric plasma. The
theory of GRMHD is the most active field of study to examine the
dynamics of magnetized plasma and impressions of black hole gravity.

The de-Sitter black hole is a vacuum solution of the Einstein field
equations with a positive cosmological constant \cite{4}. The
Schwarzschild de-Sitter (SdS) spacetime expresses a black hole
describing a patch of the de-Sitter spacetime. Inclusion of the
positive cosmological constant in the Schwarzschild spacetime leads
to SdS black hole which is non-rotating, thus plasma in
magnetosphere falls along the radial direction. Current observations
indicate that our universe is accelerating, inclusion of positive
cosmological constant leads to the expanding universe \cite{5}-\cite{7}. Thus our
universe approaches to de-Sitter universe in far future.

Regge and Wheeler \cite{8} and Zerilli \cite{9}
investigated the stability and gravitational field of the
Schwarzschild black hole with the help of non-spherical
perturbations. Petterson \cite{10} found that gravitational field is
more strong near the surface of non-rotating black hole. Price
\cite{11} explored the dynamics of nearly spherical star by
considering linear perturbation. Gleiser et al. \cite{12} discussed the
stability of black holes by applying second order perturbations.

The more appropriate approach towards General Relativity is to
decompose metric field into layers of three-dimensional spacelike
hypersurfaces and one-dimensional time, called as ADM $3+1$
formalism, presented by Arnowitt, Deser and Misner (ADM) \cite{13}.
Thorne and Macdonald \cite{14} examined that $3+1$ split is
suitable approach for black hole theory. Smarr and York \cite{15} used
ADM formalism to explain spacetime kinematics numerically. Israel
\cite{16} studied event horizons in static vacuum and static
electro-vacuum spacetimes. Zhang \cite{17} reformulated the laws of
perfect GRMHD for the general line element and discussed the
significant features of stationary symmetric GRMHD solutions. The
same author \cite{18} studied the dynamics of rotating black
hole. Buzzi et al. \cite{19} investigated wave analysis of two fluid
plasma model around Schwarzschild event horizon.

Rezolla et al. \cite{20} considered the effects of cosmological
constant and provided the dynamics of thick disks around SdS black
hole. Myung \cite{21} studied the entropy change for SdS black hole.
Suneeta \cite{22} presented the quasinormal modes for scalar field
perturbations of SdS black hole. Ali and Rahman \cite{23} investigated
transverse wave propagation in two fluid plasma for SdS black hole.
Sharif and his collaborators \cite{24}-\cite{27} found dispersion modes of cold and isothermal
plasmas for non-rotating as well as rotating black holes in the
usual medium. They have also investigated the wave properties around
Schwarzschild magnetosphere for the hot plasma.

Artificial materials which exhibit unusual electromagnetic
properties are termed as metamaterials. Veselago medium is the
well-known metamaterial named after Russian physicist Veselago
\cite{28}, which has negative electric permittivity and magnetic
permeability simultaneously. It has various names, double negative
medium (DNM), backward wave medium (BWM), left-handed medium (LHM),
negative refractive index medium (NIM) or negative phase velocity
medium (NPV). Many people \cite{29}-\cite{33} have explored the unusual properties of such a medium.
Ziolkowski and Heyman \cite{34} provided wave properties of NPV both
analytically and numerically. Ramakrishna \cite{35} discussed the
importance of NIM for the perfect lensing. Sharif and Mukhtar
\cite{36} explored wave properties around Schwarzschild
magnetosphere for isothermal as well as hot plasma in this unusual
medium.

In a recent paper (Sharif and Noureen, submitted), we have extended
this study to SDS black hole for isothermal plasma. Here we
investigate wave analysis of hot plasma for SdS black hole in NPV.
For this purpose, the $3+1$ GRMHD equations are considered to
determine dispersion relations with the help of Fourier analysis
after insertion of linear perturbation to the GRMHD equations for
both non-magnetized and magnetized backgrounds. The three
dimensional plot of wave vector, refractive index and change in
refractive index describes the wave properties of hot plasma. The
paper is arranged as follows: In section \textbf{2}, linearly
perturbed $3+1$ GRMHD equations and their Fourier analysis is
considered for hot plasma. The Fourier analyzed form of the GRMHD
equations for rotating (non-magnetized and magnetized) plasmas are
taken in sections \textbf{3} and \textbf{4} respectively to discuss
the corresponding wave properties. Last section provides the
conclusion.

\section{GRMHD Equations for Hot Plasma in a Veselogo Medium}

The general line element in ADM $3+1$ formalism is represented as
follows \cite{18}
\begin{equation}\setcounter{equation}{1}\label{1}
ds^2=-\alpha^2dt^2+\eta_{ij}(dx^i+\beta^idt)(dx^j+\beta^jdt).
\end{equation}
Here $\alpha$ denotes lapse function (ratio of FIDO proper time to
universal time i.e., $\frac{d\tau}{dt}$), a natural observer linked
with this spacetime is called fiducial observer (FIDO), $\beta^i$
describes three-dimensional shift vector and $\eta_{ij}~(i,j=1,2,3)$
denote the components of three-dimensional spacelike hypersurfaces.
The SdS spacetime in Rindler coordinates is given by \cite{23}
\begin{equation}\label{2}
ds^2=-\alpha^2(z)dt^2+dx^2+dy^2+dz^2,
\end{equation}
where the directions $z,~y$ and $x$ are analogous to the
Schwarzschild coordinates $r,~\phi$ and $\theta$ respectively. The
SdS black hole is non-rotating, thus the shift vector vanishes. From
the comparison of Eqs.(\ref{1}) and (\ref{2}), we have
\begin{equation}\label{3}
\alpha=\alpha(z),\quad\beta=0,\quad\eta_{ij}=1~(i=j).
\end{equation}

Appendix provides the $3+1$ GRMHD equations for the line element
(\ref{2}) in a Veselago medium, i.e. Eqs.(\ref{49})-(\ref{53}). The
equation of state for hot plasma is \cite{17}
\begin{eqnarray}\label{3}
\mu=\frac{\rho+p}{\rho_0},
\end{eqnarray}
where the rest mass density, moving mass density, pressure and
specific enthalpy are denoted by $\rho_0,~\rho,~p$ and $\mu$
respectively. For hot plasma, the specific enthalpy is not constant
which shows that there is energy exchange between plasma and
magnetic field of fluid. The $3+1$ GRMHD equations
((\ref{49})-(\ref{53})) for hot plasma around the SdS black hole
takes the form
\begin{eqnarray}\label{4}
&&\frac{\partial \textbf{B}}{\partial
t}=-\nabla\times(\alpha \textbf{V}\times \textbf{B}),\\
\label{5}&&\nabla.\textbf{B}=0,\\
\label{6} &&\frac{\partial (\rho+p) }{\partial t}+(\rho+p)\gamma^2
\textbf{V}. \frac{\partial \textbf{V}}{\partial t}+ (\rho+p)\gamma^2
V.(\alpha \textbf{V}.\nabla)
\textbf{V}\nonumber\\
&&+(\rho+p) \nabla.(\alpha\textbf{V})=0,\\\label{7}
&&\left\{\left((\rho+p)\gamma^2+\frac{\textbf{B}^2}{4\pi}\right)\delta_{ij}
+(\rho+p)\gamma^4V_iV_j-\frac{1}{4\pi}B_iB_j\right\}
\left(\frac{1}{\alpha}\frac{\partial}{\partial
t}\right.\nonumber\\
&&\left.+\textbf{V}.\nabla\right)V^j+\gamma^2V_i(\textbf{V}.\nabla)(\rho+p)
-\left(\frac{\textbf{B}^2}{4\pi}\delta_{ij}-\frac{1}{4\pi}B_iB_j\right)V^j_{,k}V^k\nonumber\\
&&=-(\rho+p)\gamma^2a_i-p_{,i}+\frac{1}{4\pi}
(\textbf{V}\times\textbf{B})_i\nabla.(\textbf{V}\times\textbf{B})
-\frac{1}{8\pi\alpha^2}(\alpha\textbf{B})^2_{,i}\nonumber\\
&&+\frac{1}{4\pi\alpha}(\alpha B_i)_{,j}B^j-\frac{1}{4\pi\alpha}
[\textbf{B}\times\{\textbf{V}\times(\nabla\times(\alpha\textbf{V}\times\textbf{B}))\}]_i,\\
\label{8} &&(\frac{1}{\alpha}\frac{\partial}{\partial
t}+\textbf{V}.\nabla)(\rho+p)\gamma^2-\frac{1}{\alpha}\frac
{\partial p}{\partial t}+2(\rho+p)\gamma^2(\textbf{V}.\textbf{a})
+(\rho+p)\nonumber\\
&&\gamma^2(\nabla.\textbf{V})
-\frac{1}{4\pi\alpha}\left.(\textbf{V}\times\textbf{B}).(\textbf{V}\times\frac
{\partial \textbf{B}}{\partial t}\right.)
-\frac{1}{4\pi\alpha}\left.(\textbf{V}\times\textbf{B}).(\textbf{B}\times\frac{\partial
\textbf{V}}{\partial
t}\right.)\nonumber\\&&+\frac{1}{4\pi\alpha}\left(\textbf{V}\times\textbf{B}).
(\nabla\times\alpha\textbf{B}\right.)=0.
\end{eqnarray}
We assume that for rotating background plasma flow is in two
dimensions, i.e., in $xz$-plane. Thus FIDO's measured velocity
$\textbf{V}$ and magnetic field $\textbf{B}$ become
\begin{eqnarray}\label{9}
\textbf{V}=V(z)\textbf{e}_x+u(z)\textbf{e}_z,\quad
\textbf{B}=B[\lambda(z)\textbf{e}_x+\textbf{e}_z],
\end{eqnarray}
where $B$ is an arbitrary constant. The following expression
provides the relation between the quantities $\lambda,~u$ and $V$
\cite{24}
\begin{equation}\label{a}
V=\frac{V^F}{\alpha}+\lambda u,
\end{equation}
where $V^F$ is constant of integration. The Lorentz factor,
$\gamma=\frac{1}{\sqrt{1-\textbf{V}^2}}$, takes the form
\begin{equation}\label{b}
\gamma=\frac{1}{\sqrt{1-u^2-V^2}}.
\end{equation}

Strong gravity of black hole perturbs the plasma flow. To determine
the effect of black hole gravity on plasma flow, we apply linear
perturbation. The flow variables (mass density $\rho$, pressure $p$,
velocity $\textbf{V}$ and magnetic field $\textbf{B}$) become
\begin{eqnarray}\label{10}
&&\rho=\rho^0+\delta\rho=\rho^0+\rho\widetilde{\rho},\quad
p=p^0+\delta p=p^0+p\widetilde{p},\nonumber\\
&&\textbf{V}=\textbf{V}^0+\delta\textbf{V}=\textbf{V}^0+\textbf{v},~
\textbf{B}=\textbf{B}^0+\delta\textbf{B}=\textbf{B}^0+B\textbf{b},
\end{eqnarray}
where $\rho^0,~p,~\textbf{V}^0$ and $~\textbf{B}^0$ represent the
unperturbed quantities. The linearly perturbed quantities are
denoted by $\delta\rho,~\delta p,~\delta\textbf{V}$ and
$\delta\textbf{B}$. We introduce the following dimensionless
quantities $\widetilde{\rho},~\widetilde{p},~v_x,~v_z,~b_x$ and
$b_z$ which correspond to the perturbed quantities
\begin{eqnarray}\label{11}
&&\tilde{\rho}=\tilde{\rho}(t,z),\quad
\tilde{p}=\tilde{p}(t,z),\quad\textbf{v}=\delta\textbf{V}=v_x(t,z)\textbf{e}_x
+v_z(t,z)\textbf{e}_z,\nonumber\\
&&\textbf{b}=\frac{\delta\textbf{B}}{B}=b_x(t,z)\textbf{e}_x
+b_z(t,z)\textbf{e}_z.
\end{eqnarray}
After the insertion of linear perturbations in the perfect GRMHD
equations (Eqs.(\ref{4})-(\ref{8})) and using Eq.(\ref{11}), the
component form of linearly perturbed GRMHD equations become \cite{36}
\begin{eqnarray}\label{17}
&&\frac{1}{\alpha}\frac{\partial b_x}{\partial
t}-ub_{x,z}=(ub_x-Vb_z-v_x+\lambda v_z)\nabla
\ln\alpha\nonumber\\
&&-(v_{x,z}-\lambda
v_{z,z}-\lambda'v_z+V'b_z+Vb_{z,z}-u'b_x),\\\label{18}
&&\frac{1}{\alpha}\frac{\partial b_z}{\partial t}=0,\\\label{19}
&&b_{z,z}=0,
\end{eqnarray}
\begin{eqnarray}\label{20}
&&\rho\frac{1}{\alpha}\frac{\partial\tilde{\rho}}{\partial t}
+p\frac{1}{\alpha}\frac{\partial\tilde{p}}{\partial
t}+(\rho+p)\gamma^2V(\frac{1}{\alpha}\frac{\partial{v_x}}{\partial
t}+uv_{x,z})+(\rho+p)\gamma^2u\nonumber\\
&&\times\frac{1}{\alpha}\frac{\partial{v_z}}{\partial
t}+(\rho+p)(1+\gamma^2u^2)v_{z,z}=-\gamma^2u(\rho+p)[(1+2\gamma^2V^2)V'\nonumber\\
&&+2\gamma^2uVu']v_x+(\rho+p)[(1-2\gamma^2u^2)(1+\gamma^2u^2)\frac{u'}{u}\nonumber\\
&&-2\gamma^4u^2VV']v_z,\\ \label{21}
&&\left\{(\rho+p)\gamma^2(1+\gamma^2V^2)
+\frac{B^2}{4\pi}\right\}\frac{1}{\alpha}\frac{\partial
v_x}{\partial t}+\left\{(\rho+p)\gamma^4uV-\frac{\lambda B^2}{4\pi}\right\}\nonumber\\
&&\times\frac{1}{\alpha}\frac{\partial v_z}{\partial
t}+\left\{(\rho+p)\gamma^2(1+\gamma^2V^2)
+\frac{B^2}{4\pi}\right\}uv_{x,z}+\left\{(\rho+p)\gamma^4uV\right.\nonumber\\
&&\left.-\frac{\lambda B^2}{4\pi}\right\}uv_{z,z}
-\frac{B^2}{4\pi}(1+u^2)b_{x,z}-\frac{B^2}{4\pi\alpha}\left\{\alpha'(1+u^2)+\alpha
uu'\right\}b_x\nonumber\\
&&+\gamma^2u(\rho\tilde{\rho}+p\tilde{p})\left\{(1+\gamma^2V^2)V'+\gamma^2uVu'\right\}
+\gamma^2uV(\rho'\tilde{\rho}+\rho\tilde{\rho}'\nonumber\\
&&+p'\tilde{p}+p\tilde{p}')+[(\rho+p)\gamma^4u
\left\{(1+4\gamma^2V^2)uu'+4VV'(1+\gamma^2V^2)\right\}\nonumber\\
&&+\frac{B^2u\alpha'}{4\pi\alpha}
+\gamma^2u(1+2\gamma^2V^2)(\rho'+p')]v_x
+[(\rho+p)\gamma^2\left\{(1+2\gamma^2u^2)\right.\nonumber\\
&&\left.(1+2\gamma^2V^2)V'-\gamma^2V^2V'
+2\gamma^2(1+2\gamma^2u^2)uVu'\right\}-\frac{B^2u}
{4\pi\alpha}(\lambda\alpha)'\nonumber\\
&&+\gamma^2V(1+2\gamma^2u^2)(\rho'+p')]v_z=0, \\\label{22}
&&\left\{(\rho+p)\gamma^2(1+\gamma^2u^2)
+\frac{\lambda^2B^2}{4\pi}\right\}\frac{1}{\alpha}\frac{\partial
v_z}{\partial t}+\left\{(\rho+p)\gamma^4uV -\frac{\lambda B
^2}{4\pi}\right\}\nonumber\\
&&\times\frac{1}{\alpha}\frac{\partial v_x}{\partial t}
+\left\{(\rho+p)\gamma^2(1+\gamma^2u^2)+\frac{\lambda^2B^2}{4\pi}\right\}
uv_{z,z}+\left\{(\rho+p)\gamma^4uV \right.\nonumber\\
&&\left.-\frac{\lambda B^2}{4\pi}\right\}uv_{x,z}+\frac{\lambda
B^2}{4\pi}(1+u^2)b_{x,z}+\frac{B^2}{4\pi\alpha}\left\{(\alpha\lambda)'
-\alpha'\lambda+u\lambda(u\alpha'\right.\nonumber\\
&&\left.+u'\alpha)\right\}b_x+(\rho\tilde{\rho}+p\tilde{p})\gamma^2\left\{a_z
+uu'(1+\gamma^2u^2)+\gamma^2u^2VV'\right\}\nonumber\\
&&+(1+\gamma^2u^2)(p'\tilde{p}+p\tilde{p}')
+\gamma^2u^2(\rho'\tilde{\rho}+\rho\tilde{\rho}')+[(\rho+p)\gamma^4\nonumber\\
&&\times\{u^2V'(1+4\gamma^2V^2)+2V(a_z+uu'(1+2\gamma^2u^2))\}-\frac{\lambda
B^2u\alpha'}{4\pi\alpha}\nonumber\\
&&+2\gamma^4u^2V(\rho'+p')]v_x+[(\rho+p)\gamma^2
\left\{u'(1+\gamma^2u^2)(1+4\gamma^2u^2)\right.\nonumber\\
&&\left.+2u\gamma^2(a_z+(1+2\gamma^2u^2)VV')\right\} +\frac{\lambda
B^2u}{4\pi\alpha}(\alpha\lambda)'
+2\gamma^2u(1\nonumber\\
&&+\gamma^2u^2)(\rho'+p')]v_z=0,
\end{eqnarray}
\begin{eqnarray}\label{23}
&&\frac{1}{\alpha}\gamma^2\rho\frac{\partial \tilde{\rho}}{\partial
t}+\frac{1}{\alpha}\gamma^2p\frac{\partial \tilde{p}}{\partial
t}+\gamma^2(\rho'+p')v_z+u\gamma^2(\rho\tilde{\rho}_{,z}
+p\tilde{p}_{,z}+\rho'\tilde{\rho}+p'\tilde{p})\nonumber\\&&
-\frac{1}{\alpha}p\frac{\partial \tilde{p}}{\partial
t}+2\gamma^2u(\rho\tilde{\rho}+p\tilde{p})a_z
+\gamma^2u'(\rho\tilde{\rho}+p\tilde{p})+2(\rho+p)\gamma^4(uV'\nonumber\\
&&+2uVa_z+u'V)v_x+2(\rho+p)\gamma^2(2\gamma^2uu'+a_z\gamma^4+2\gamma^2u^2a_z)v_z+\nonumber\\
&&2(\rho+p)\gamma^4uVv_{x,z}+(\rho+p)\gamma^2(1+2\gamma^2u^2)v_{z,z}
-\frac{B^2}{4\pi\alpha}[(V^2+u^2)\lambda\frac{\partial b_x}{\partial
t}\nonumber\\
&&+(V^2+u^2)\frac{\partial b_z}{\partial t}-\lambda
V(\lambda V+u)\frac{\partial b_x}{\partial t}-u(\lambda
V+u)\frac{\partial
b_z}{\partial t}]-\frac{B^2}{4\pi\alpha}\nonumber\\
&&\times[\lambda(\lambda V+u)v_{x,t}+(\lambda V+u)v_{z,t}
-(\lambda^{2}+1)Vv_{x,t}-(\lambda^{2}+1)uv_{z,t}]\nonumber\\
&&+\frac{B^2}
{4\pi}(\lambda\lambda'v_z-\lambda'v_x-\lambda'Vb_z+\lambda'ub_x-V
b_{x,z}+u\lambda b_{x,z})=0.
\end{eqnarray}

For the Fourier analysis of the linearly perturbed GRMHD equations,
we take the following harmonic spacetime dependence
\begin{eqnarray}\label{24}
\widetilde{\rho}(t,z)=c_1e^{-\iota(\omega t-kz)},&\quad&
\widetilde{p}(t,z)=c_2e^{-\iota(\omega t-kz)},\nonumber\\
v_z(t,z)=c_3e^{-\iota(\omega t-kz)},&\quad&
v_x(t,z)=c_4e^{-\iota(\omega t-kz)},\nonumber\\
b_z(t,z)=c_5e^{-\iota(\omega t-kz)},&\quad&
b_x(t,z)=c_6e^{-\iota(\omega t-kz)},
\end{eqnarray}
where $\omega$ represents angular frequency and $k$ denotes the
$z$-component of the wave vector $(0,0,k)$. The wave vector is
helpful to determine refractive index and the properties of plasma
about the event horizon. Frequency dependence effects on the wave
propagating in a medium are related to dispersion. Using
Eq.(\ref{24}) in Eqs.(\ref{17})-(\ref{23}), we obtain their Fourier
analyzed form
\begin{eqnarray}\label{25}
&&c_{4}(\alpha'+\iota k\alpha)-c_3\ \left\{(\alpha\lambda)'+\iota
k\alpha\lambda\ \right\}-c_5(\alpha V)'-c_6\{(\alpha
u)'-\iota\omega\nonumber\\
&&+\iota ku\alpha\}=0,\\\label{26}
&&c_5(\frac{-\iota\omega}{\alpha})=0,\\\label{27} &&c_5\iota
k=0,\\\label{28}&&c_1(\frac{-\iota\omega}{\alpha}\rho)+c_2(\frac{-\iota\omega}{\alpha}p)
+c_3(\rho+p)[\frac{-\iota\omega}{\alpha}\gamma^2u+(1+\gamma^2u^2)\iota k\nonumber\\
&&-(1-2\gamma^2u^2)(1+\gamma^2u^2)\frac{u'}{u}+2\gamma^4u^2VV']
+c_4(\rho+p)\gamma^2[(\frac{-\iota\omega}{\alpha}\nonumber\\
&&+\iota ku)V+u(1+2\gamma^2V^2)V'+2\gamma^2u^2Vu']=0,
\end{eqnarray}
\begin{eqnarray}\label{29}
&&c_1[\rho\gamma^2u\{(1+\gamma^2V^2)V'+\gamma^2Vuu'\}+\gamma^2Vu(\rho'+\iota
k\rho)]\nonumber\\
&&+c_2[p\gamma^2u\{(1+\gamma^2V^2)V'+\gamma^2Vuu'\}+\gamma^2Vu(p'+\iota kp)]\nonumber\\
&&+c_3[(\rho+p)\gamma^2
\{(1+2\gamma^2u^2)(1+2\gamma^2V^2)V'+(\frac{-\iota\omega}{\alpha}+\iota
ku)\gamma^2Vu\nonumber\\
&&-\gamma^2V^2V'+2\gamma^2(1+2\gamma^2u^2)uVu'\}+\gamma^2V(1+2\gamma^2u^2)(\rho'+p')\nonumber\\
&&-\frac{B^2u}{4\pi\alpha}(\lambda\alpha)'+\frac{\lambda
B^2}{4\pi}(\frac{\iota\omega}{\alpha}-\iota
ku)]+c_4[(\rho+p)\gamma^4u\{(1+4\gamma^2V^2)\nonumber\\ &&\times
uu'+4VV'(1+\gamma^2V^2)\}
+(\rho+p)\gamma^2(1+\gamma^2V^2)(\frac{-\iota\omega}{\alpha}+\iota
ku)\nonumber\\
&&+\gamma^2u(1+2\gamma^2V^2)(\rho'+p')+\frac{B^2u\alpha'}{4\pi\alpha}
-\frac{B^2}{4\pi}(\frac{\iota\omega}{\alpha}-\iota
ku)]\nonumber\\
&&-c_6\frac{B^2}{4\pi\alpha}[\alpha uu'+\alpha'(1+u^2)+(1+u^2)\iota
k\alpha]=0, \\\label{30}
&&c_1[\rho\gamma^2\{a_z+(1+\gamma^2u^2)uu'+\gamma^2u^2VV'\}+\gamma^2u^2(\rho'+\iota
k\rho)]\nonumber\\
&&+c_2[p\gamma^2\{a_z+(1+\gamma^2u^2)uu'+\gamma^2u^2VV'\}+(1+\gamma^2u^2)\nonumber\\
&&\times(p'+\iota kp)]+c_3[(\rho+p)\gamma^2
\{(1+\gamma^2u^2)(\frac{-\iota\omega}{\alpha}+\iota ku)\nonumber\\
&&+u'(1+\gamma^2u^2)(1+4\gamma^2u^2)+2u\gamma^2(a_z
+(1+2\gamma^2u^2)VV')\}\nonumber\\
&&+2\gamma^2u(1+\gamma^2u^2)(\rho'+p')+\frac{\lambda
B^2u}{4\pi\alpha}(\lambda\alpha)'-\frac{\lambda^2
B^2}{4\pi}(\frac{\iota\omega}{\alpha}-\iota ku)]\nonumber\\
&&+c_4[(\rho+p)\gamma^4\{(\frac{-\iota\omega}{\alpha}+\iota
ku)uV+u^2V'(1+4\gamma^2V^2)+2V(a_z\nonumber\\
&&+(1+2\gamma^2u^2)uu')\}+2\gamma^4u^2V(\rho'+p')+\frac{\lambda
B^2}{4\pi}(\frac{\iota\omega}{\alpha}-\iota ku)\nonumber\\
&&-\frac{\lambda B^2u\alpha'}{4\pi\alpha}]
+c_6[\frac{B^2}{4\pi\alpha}\{-(\lambda\alpha)'
+\alpha'\lambda-u\lambda(u\alpha'+u'\alpha)\}\nonumber\\
&&+\frac{\lambda B^2}{4\pi}(1+u^2)\iota k]=0,\\
\label{31}&&c_1\{(\frac{-\iota\omega}{\alpha}\gamma^2+\iota ku
\gamma^2+2u\gamma^2a_z+\gamma^2u')\rho+u\rho'\gamma^2\}
+c_2\{(\frac{\iota\omega}{\alpha}(1-\gamma^2)\nonumber\\
&&+\iota ku\gamma^2+2\gamma^2ua_z+\gamma^2u')p+u\gamma^2p'\}+c_3\gamma^2\{(\rho'+p')+2\nonumber\\
&&\times(2\gamma^4uu'+a_z+2\gamma^2u^2a_z)(\rho+p)+(1+2\gamma^2u^2)(\rho+p)\iota
k+\frac{\lambda B^2}{4\pi\alpha}\nonumber\\
&&\times(\lambda
u-V)\iota\omega+\alpha\lambda'\}+c_4[2(\rho+p)\gamma^2\{(u
V'+2uVa_z+u'V)+uV\iota
k\}\nonumber\\
&&+\frac{B^2}{4\pi\alpha}(V-u\lambda)\iota\omega-\alpha\lambda']
+c_6[\frac{B^2}{4\pi\alpha}\{(\iota\omega u+\iota\alpha k)(\lambda
u-V)-\alpha\lambda'\nonumber\\&&+uV\iota\alpha k\}]=0.
\end{eqnarray}

\section{Rotating Non-Magnetized Background}

In the rotating non-magnetized plasma flow, we take $B=0=\lambda$
and $c_5=0=c_6$ in the Fourier analyzed perturbed GRMHD equations
((\ref{28})-(\ref{31})) \cite{36}.

\subsection{Numerical Solutions}

To determine numerical solutions of the Fourier analyzed equations
in rotating non-magnetized background, we assume
\begin{itemize}
\item Specific enthalpy: $\mu=\sqrt{\frac{1+\alpha^{2}}{2}}$,
\item Time lapse: $\alpha=\frac{z}{2r_h}$,
\item  Velocity components: $u=V,~x$ and $z$-components of velocity
yield $u=V=-\frac{1}{\sqrt{z^{2}+2}}$,
\item Stiff fluid: $\rho=p=\frac{\mu}{2}$,
\item Lorentz factor: $ \gamma=\frac{1}{\sqrt{1-u^2-V^2}}=
\frac{\sqrt{z^{2}+2}}{z}$,
\end{itemize}
where $r_h$ is the SdS event horizon greater than that of the
Schwarzschild event horizon and $r_h\thickapprox
2M\left(1+\frac{4M^{2}}{l^{2}}+...\right)\backsimeq\zeta 2.948km$,
$1\leqslant\zeta\leqslant1.5$ \cite{23}. It is known
that strong gravity of black hole plays an important role to perturb
plasma present in its surroundings which causes the plasma wave
propagation around black holes. The frequency dependence effects in
wave propagation leads to the dispersion, i.e., the phenomenon in
which phase or group velocity of the wave depends upon the frequency
\cite{38}.

We consider the region $4\leq z\leq10$ for wave analysis considering
that event horizon is exactly at $z=0$. The determinant of the
coefficients of constants of the corresponding equations for the
rotating non-magnetized plasma yields a complex dispersion relation.
The real part of the determinant gives a quartic equation in $k$
\begin{equation}\label{36}
A_1(z)k^4+A_2(z,\omega)k^3+A_3(z,\omega)k^2+A_4(z,\omega)k+A_5(z,\omega)=0
\end{equation}
which yields two real and two imaginary roots. A cubic equation in
$k$ is obtained from the imaginary part of dispersion relation
\begin{equation}\label{37}
B_1(z)k^3+B_2(z,\omega)k^2+B_3(z,\omega)k+B_4(z,\omega)=0
\end{equation}
which provides three real roots. The wave vector, refractive index,
its change with respect to angular frequency, group velocity and
phase velocity lead to the wave properties of the SdS black hole in
the presence of Veselago medium. These are shown by graphical
representation in Figures \textbf{1-5} expressing real values of $k$
in Eqs.(\ref{36}) and (\ref{37}).

The dispersion is normal if phase velocity is greater than the group
velocity, otherwise anomalous \cite{38} or equivalently
dispersion is normal if change in refractive index with respect to
angular frequency is positive and anomalous otherwise. The
information about energy exchange within the magnetosphere can be
extracted only where the dispersion is normal. It is clear from
figures that some waves move towards the event horizon and some move
away from the horizon. The dispersion is normal in Figures
\textbf{2}, \textbf{3} and \textbf{5}, while Figures \textbf{1} and
\textbf{4} indicate normal as well as anomalous at random points of
the region. The wave properties of non-magnetized hot plasma are
illustrated in the tables I and II.
\begin{center}
Table I. Direction and refractive index of waves
\end{center}
\begin{tabular}{|c|c|c|c|c|}
\hline\textbf{Fig.} & \textbf{Direction of Waves} &
\textbf{Refractive Index} ($n$)\\ \hline
& & $n<1$ and decreases in the region \\
\textbf{1} & Move towards the event horizon & $4\leq z\leq
9,0\leq\omega\leq10$\\&& with the decrease in $z$  \\
\hline
& & $n<1$ and increases in the region\\
\textbf{2} & Move away from the event horizon & $4\leq z\leq 9.8,
0\leq\omega\leq
10$\\&&with the decrease in $z$  \\
\hline
& & $n<1$ and increases in the region\\
\textbf{3} & Move towards the event horizon & $4\leq z\leq5.8,0\leq\omega\leq 10$\\
& &with the decrease in $z$ \\
\hline
& & $n<1$ and decreases in the region\\
\textbf{4} & Move towards the event horizon & $4\leq z\leq8.8,0\leq\omega\leq 10$\\
& &with the decrease in $z$ \\
\hline
& & $n<1$ and increases in the region\\
\textbf{5} & Moves away from the event horizon & $4\leq z\leq5.8,0\leq\omega\leq 10$\\
& &with the decrease in $z$ \\
\hline
\end{tabular}\\\\\\
The regions for anomalous and normal dispersion for figures
exhibiting random dispersion are separated as follows:
\begin{center}
\par\noindent
Table II. Regions of normal and anomalous dispersion
\end{center}
\begin{center}
\begin{tabular}{|c|c|c|c|c|}
\hline \textbf{Fig.}&  \textbf{ Normal dispersion} &
\textbf{Anomalous dispersion}
\\\hline \textbf{1} & $4.6\leq z\leq 10, 0\leq\omega\leq 10$   &
$4\leq z\leq 4.5, 0.1\leq\omega\leq 2.1$
\\\hline
\textbf{4} & $4.7\leq z\leq 10, 1\leq\omega\leq 10$   & $4\leq z\leq
4.5, 0\leq\omega\leq 3.5$
\\\hline
\end{tabular}
\end{center}
\begin{figure}
\begin{tabular}{cc}
\epsfig{file=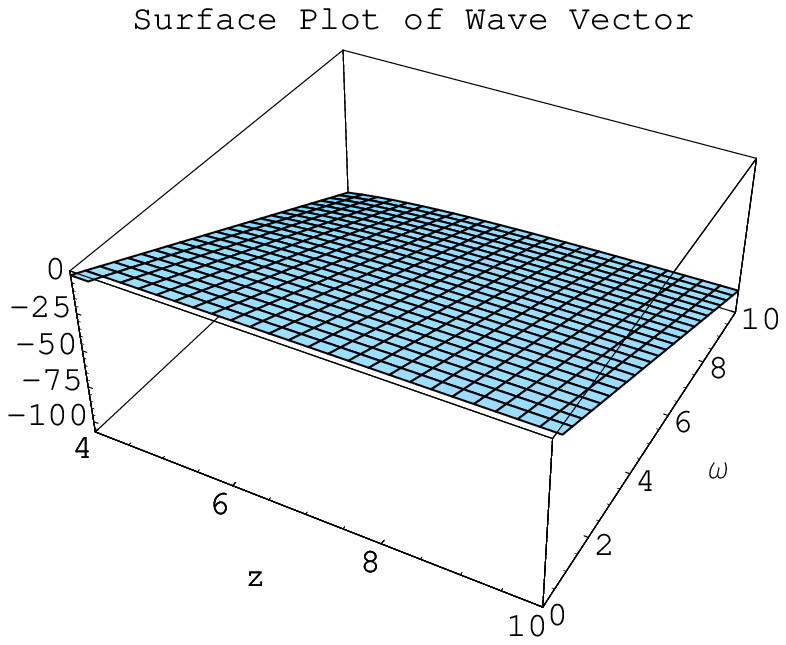,width=0.34\linewidth}
\epsfig{file=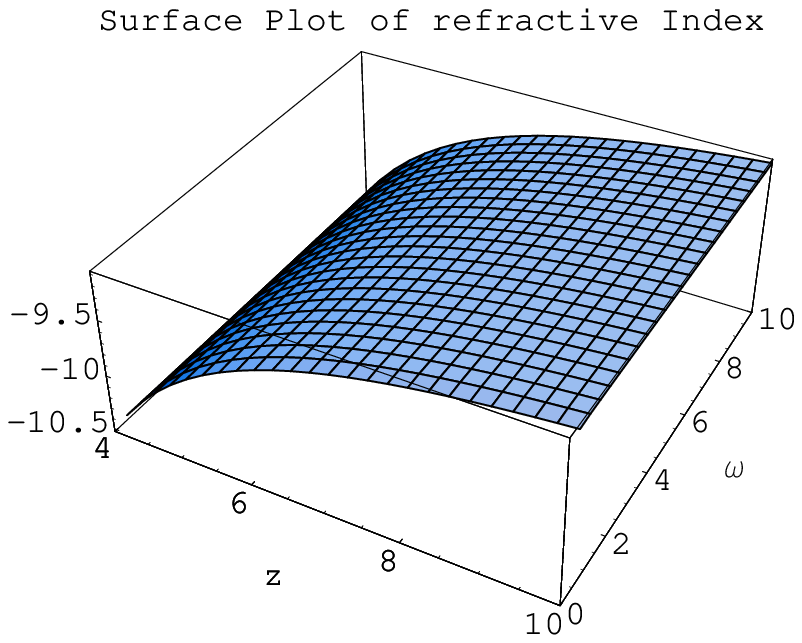,width=0.34\linewidth}\\
\epsfig{file=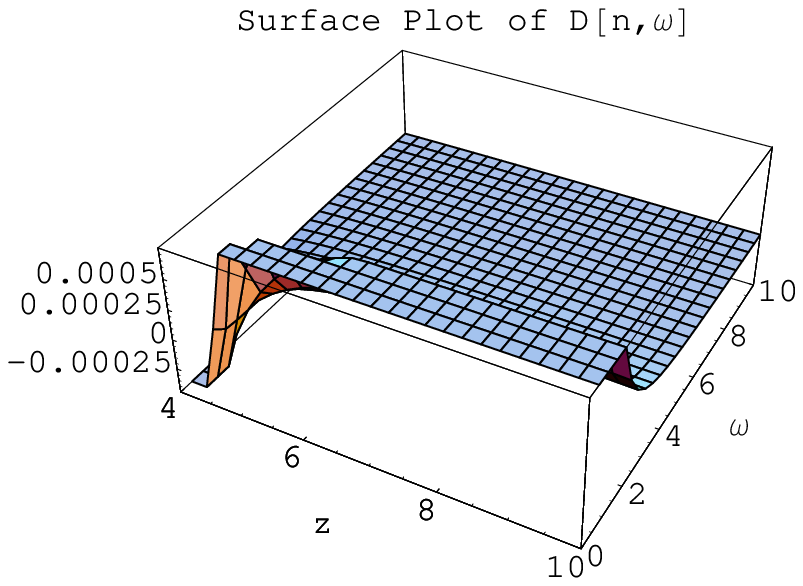,width=0.34\linewidth}
\epsfig{file=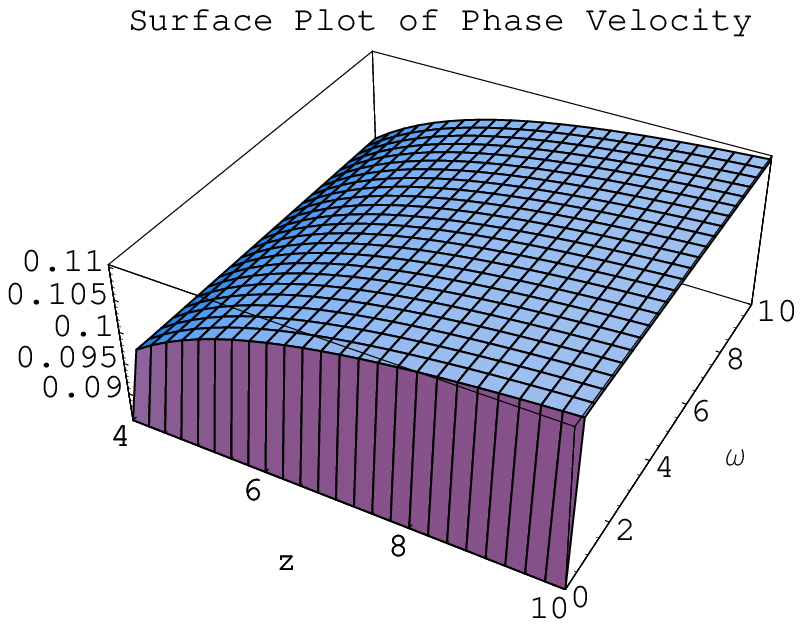,width=0.34\linewidth}
\epsfig{file=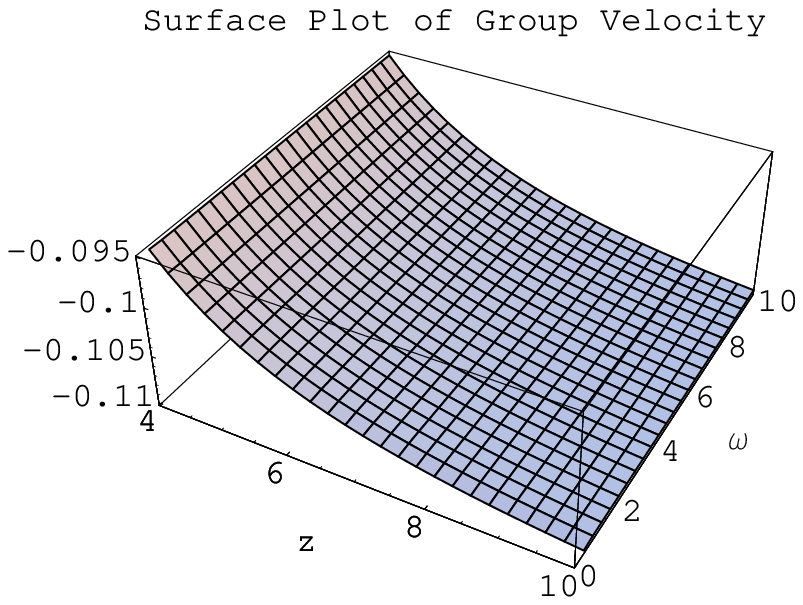,width=0.34\linewidth}\\
\end{tabular}
\caption{Dispersion is normal and anomalous in the region.}
\begin{tabular}{cc}\\
\epsfig{file=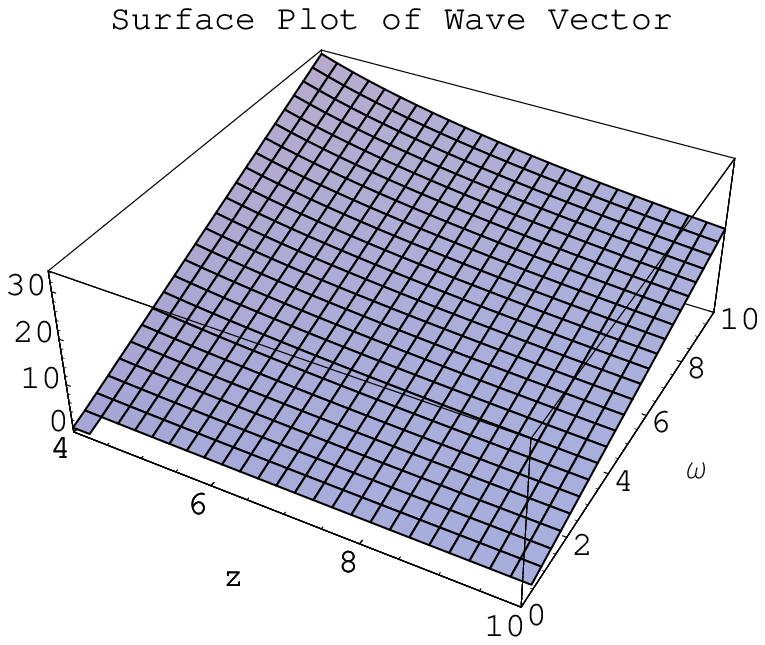,width=0.34\linewidth}
\epsfig{file=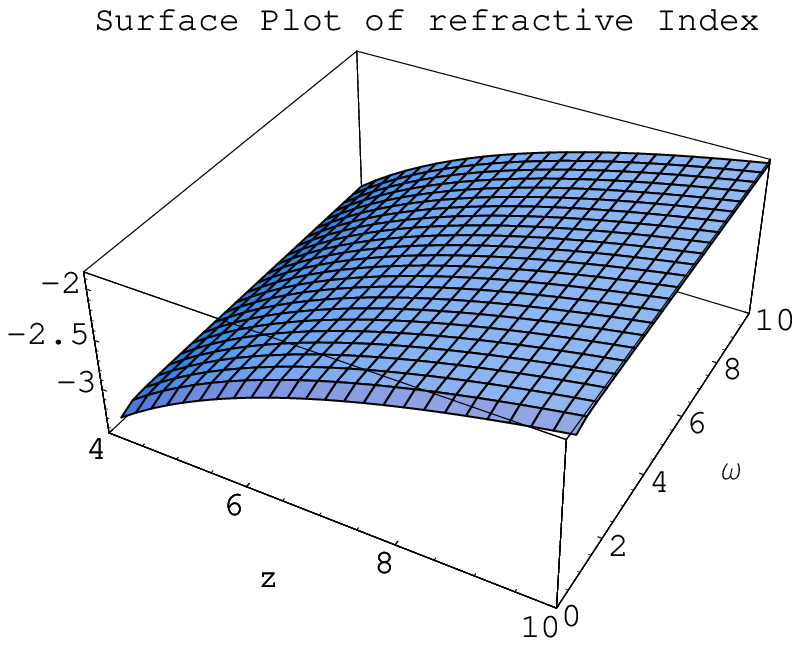,width=0.34\linewidth}\\
\epsfig{file=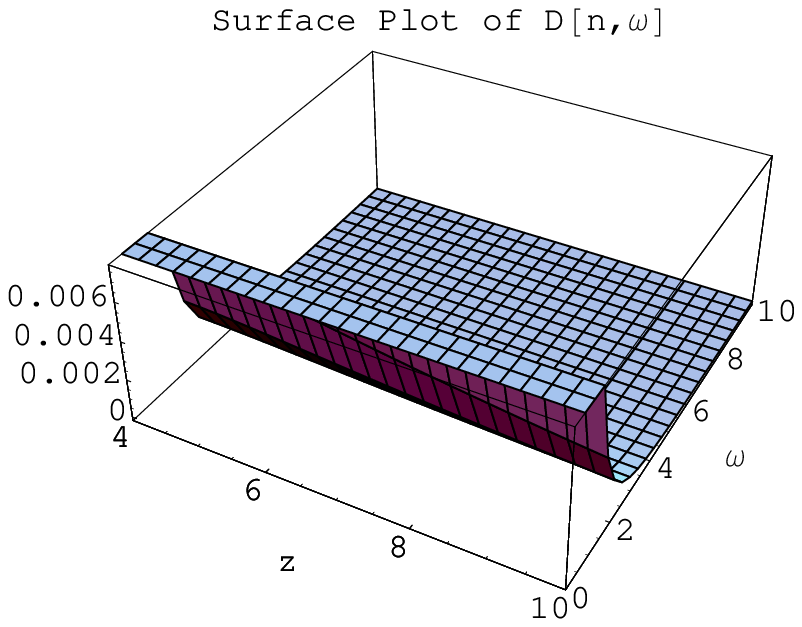,width=0.34\linewidth}
\epsfig{file=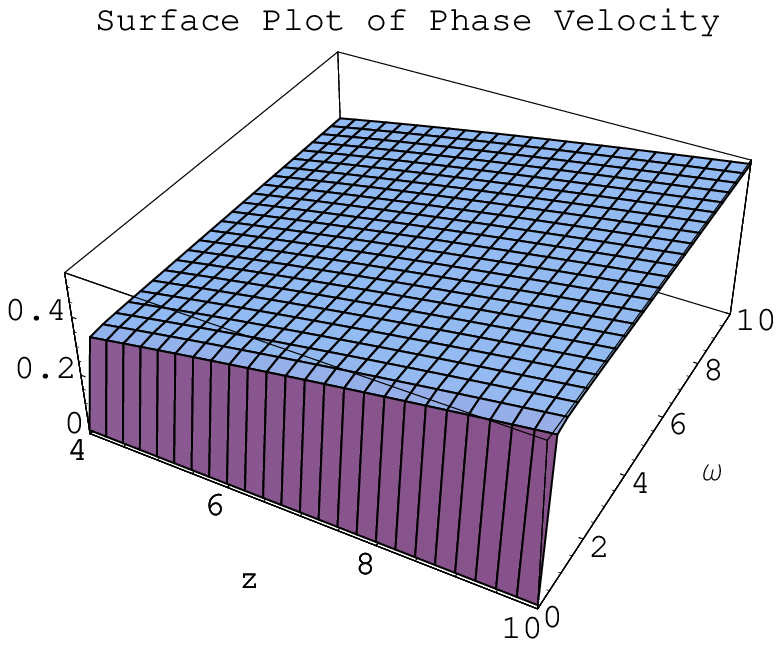,width=0.34\linewidth}
\epsfig{file=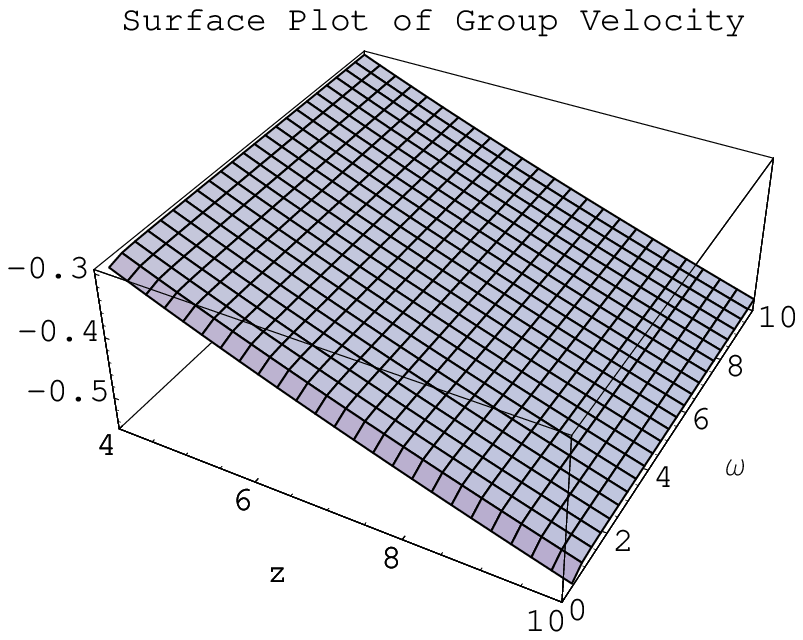,width=0.34\linewidth}\\
\end{tabular}
\caption{Whole region exhibits the normal dispersion.}
\end{figure}
\begin{figure}
\begin{tabular}{cc}
\epsfig{file=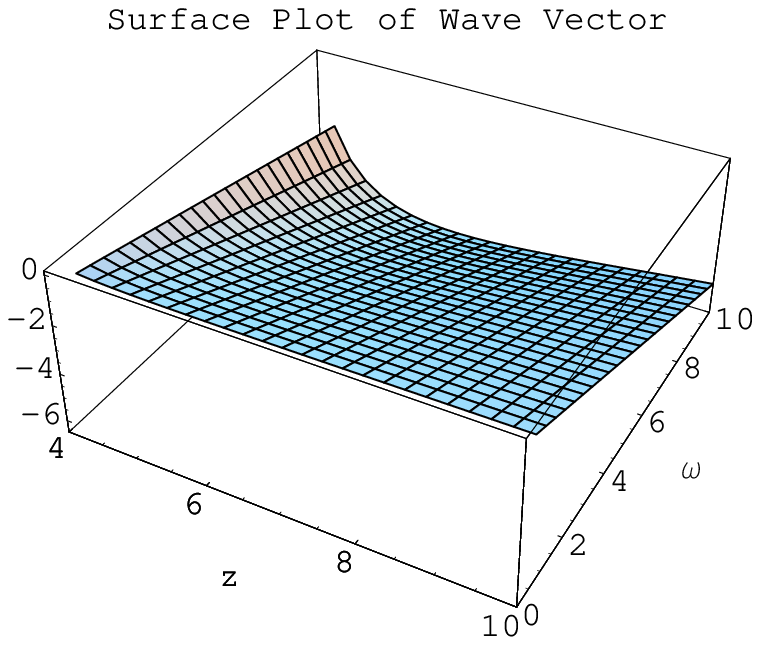,width=0.34\linewidth}
\epsfig{file=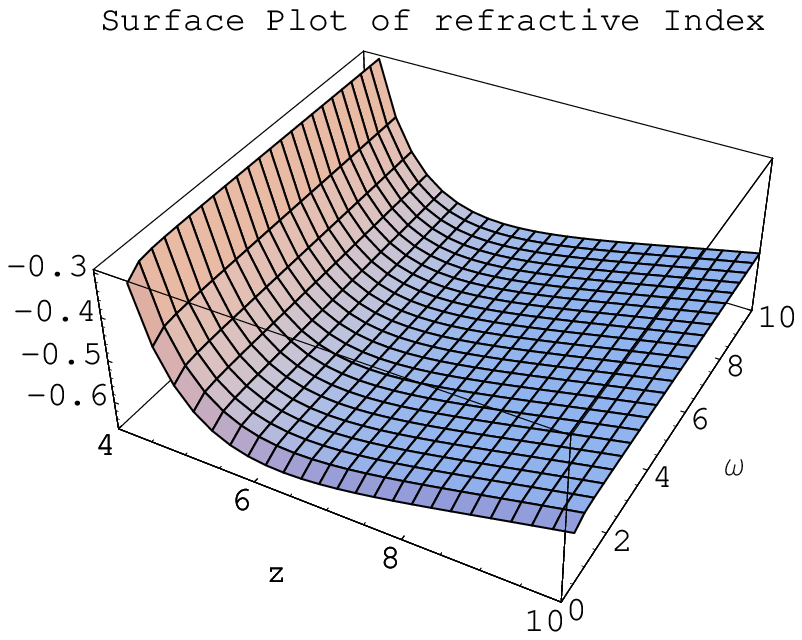,width=0.34\linewidth}\\
\epsfig{file=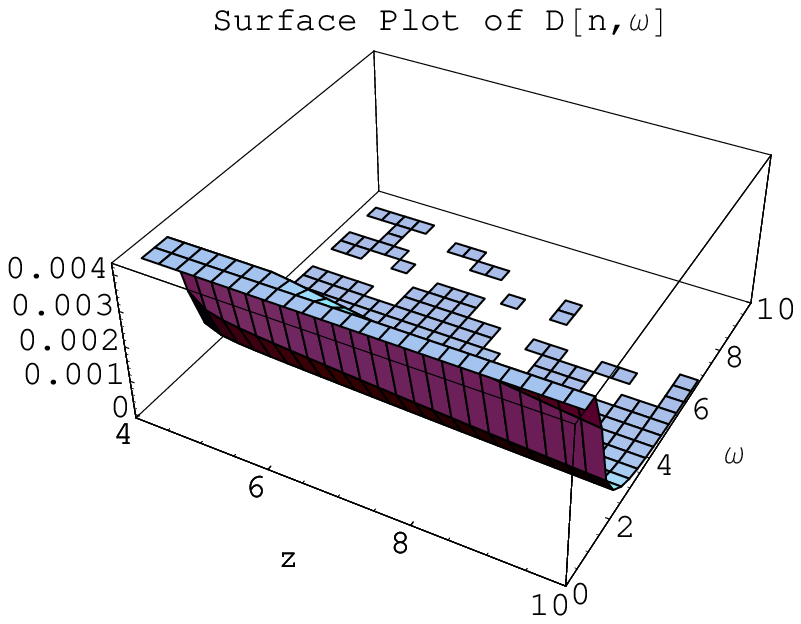,width=0.34\linewidth}
\epsfig{file=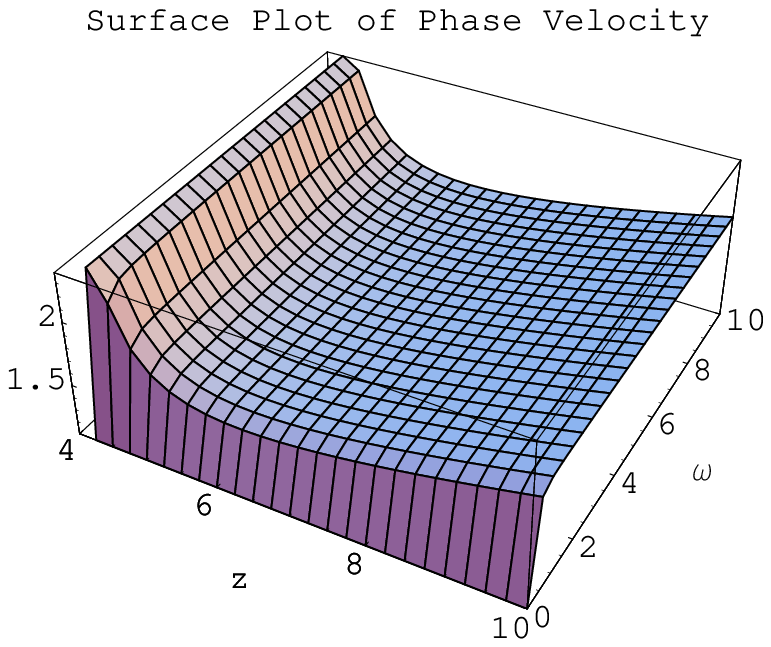,width=0.34\linewidth}
\epsfig{file=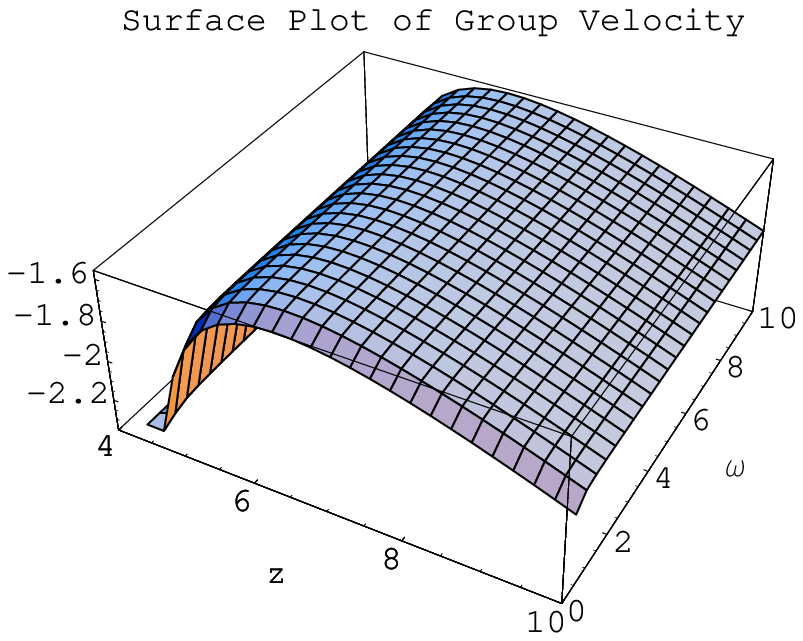,width=0.34\linewidth}\\
\end{tabular}
\caption{Normal dispersion of waves is observed.}
\begin{tabular}{cc}\\
\epsfig{file=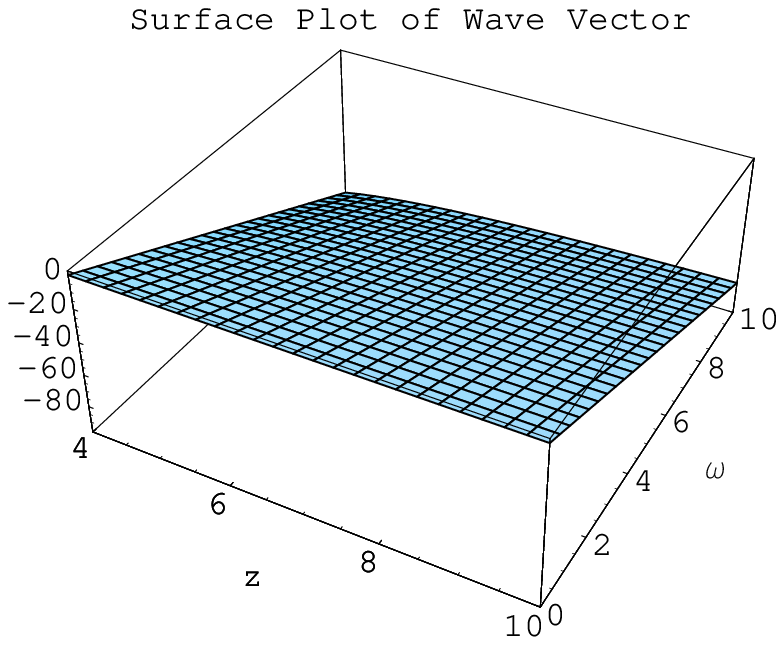,width=0.34\linewidth}
\epsfig{file=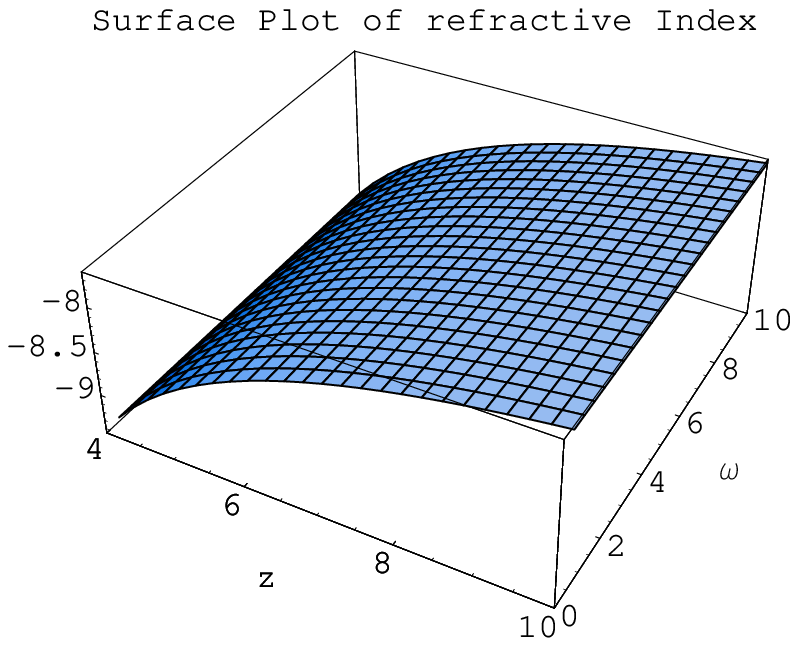,width=0.34\linewidth}\\
\epsfig{file=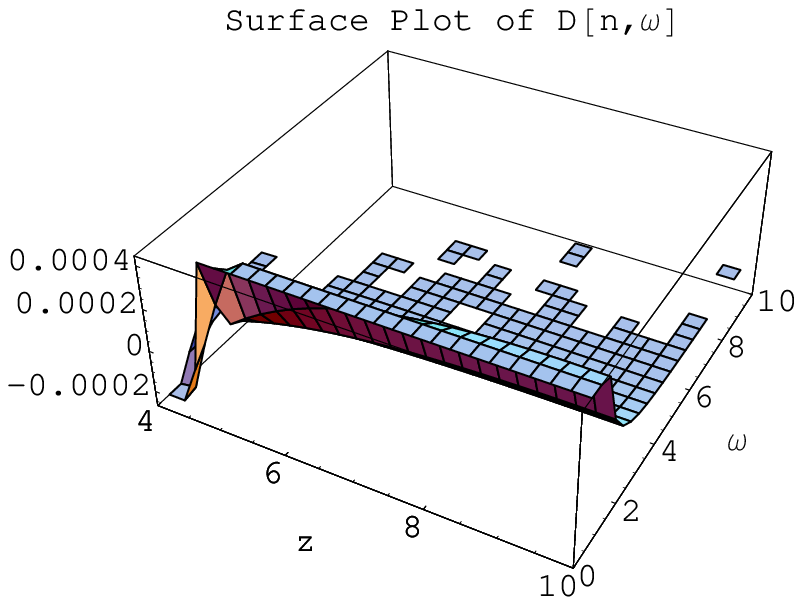,width=0.34\linewidth}
\epsfig{file=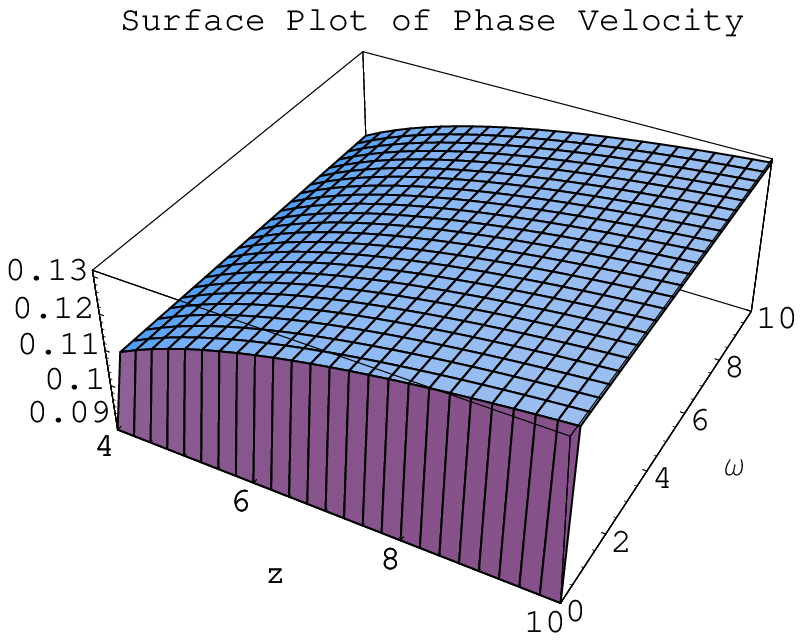,width=0.34\linewidth}
\epsfig{file=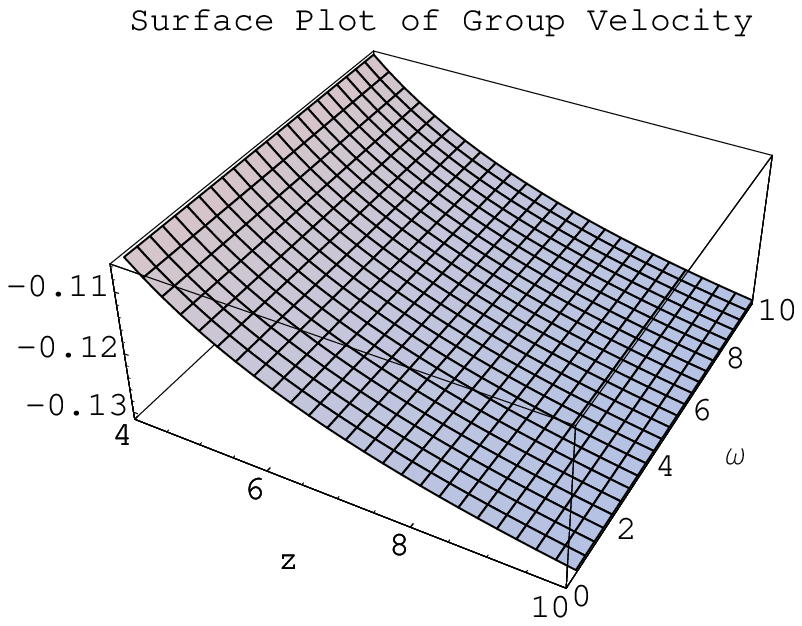,width=0.34\linewidth}\\
\end{tabular}
\caption{Random points of normal and anomalous dispersion are
found in the region.}
\end{figure}
\begin{figure}
\begin{tabular}{cc}
\epsfig{file=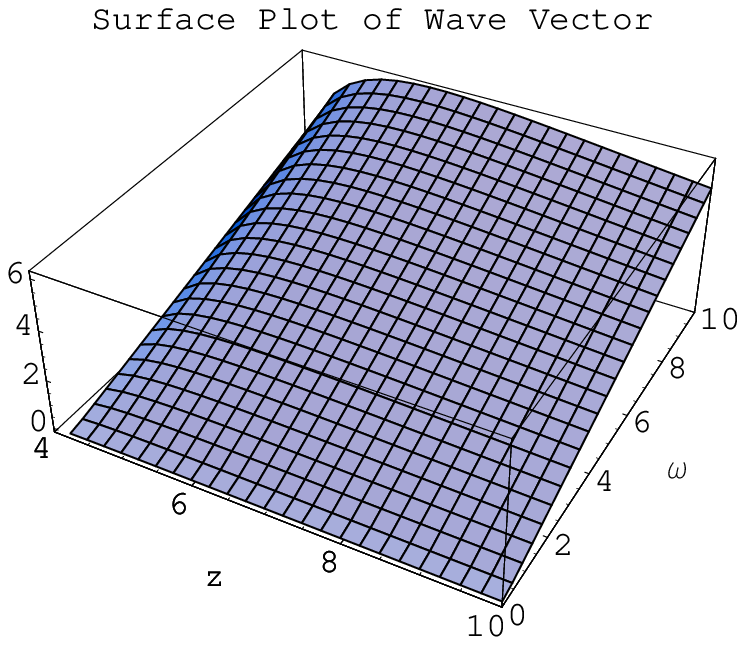,width=0.34\linewidth}
\epsfig{file=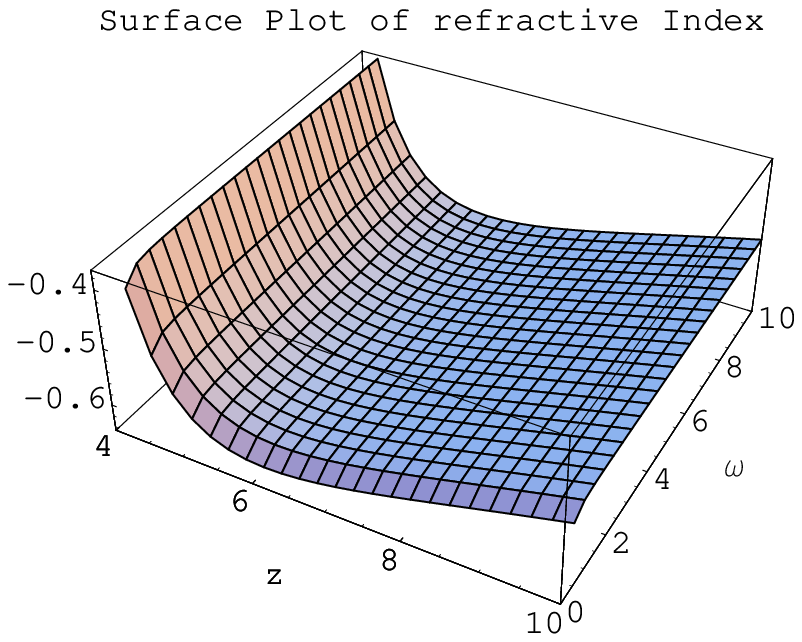,width=0.34\linewidth}\\
\epsfig{file=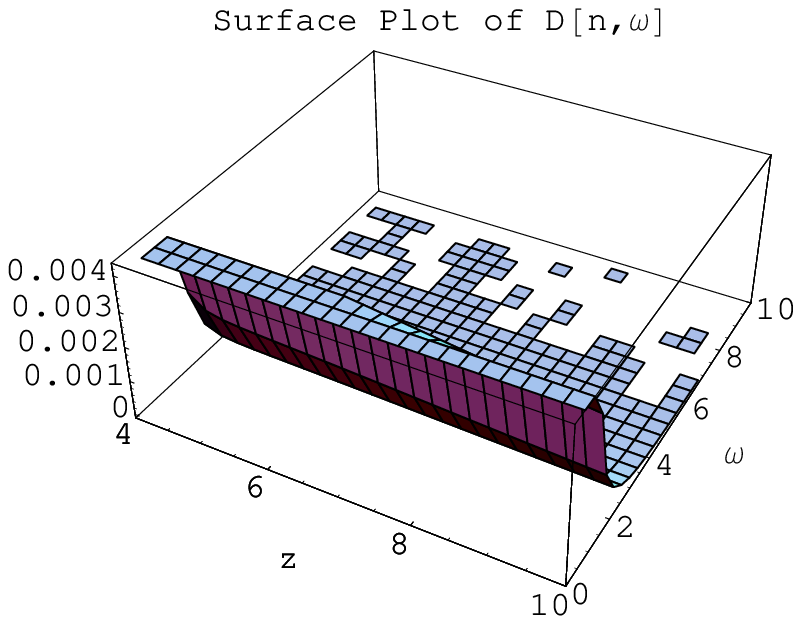,width=0.34\linewidth}
\epsfig{file=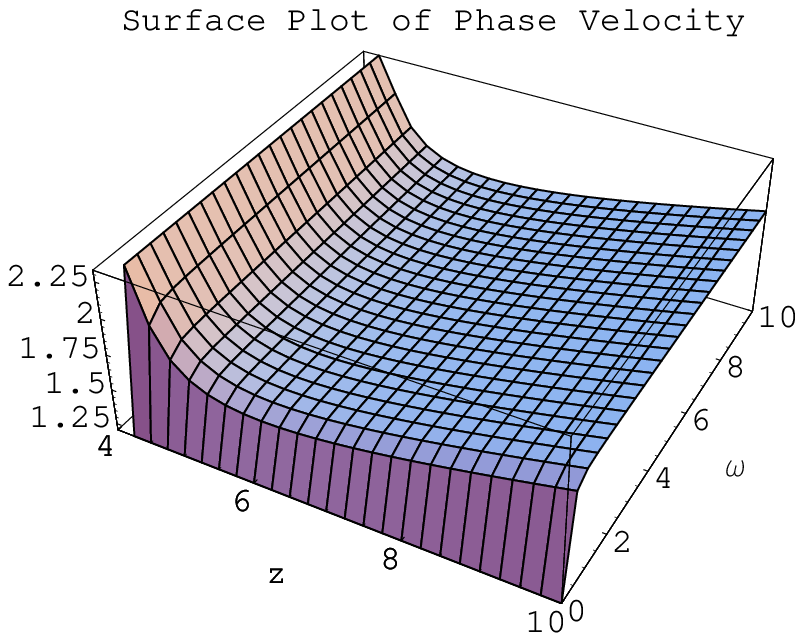,width=0.34\linewidth}
\epsfig{file=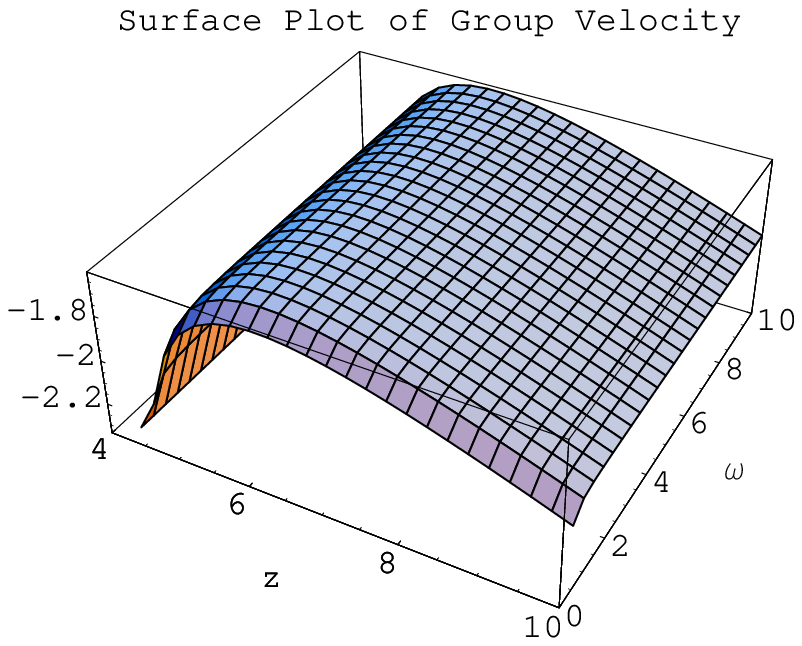,width=0.34\linewidth}\\
\end{tabular}
\caption{Whole region admits normal dispersion.}
\end{figure}

\section{Plasma Flow With Rotating Magnetized Background}

Here plasma is considered to be rotating and magnetized. The flow is
supposed to be in two dimensions, so the magnetic field and velocity
of fluid lie in $xz$-plane. Equations (\ref{25})-(\ref{31})
represent the perturbed Fourier analyzed GRMHD equations for the
rotating magnetized plasma.

\subsection{Numerical Solutions}

The assumptions for velocity, lapse function and specific enthalpy
are the same as in the previous section. We put the following
restrictions on magnetic field to find numerical solutions.
\begin{itemize}
\item $\frac{B^{2}}{4\pi}=2$ with $u=V$
\item putting $V^F=1$ in Eq.(\ref{a}) so that
$\lambda=1+\frac{\sqrt{2+z^{2}}}{z}$.
\end{itemize}
Again our region of consideration is $4\leq z\leq10$ and
$0\leq\omega\leq 10$. Equations (\ref{26})-(\ref{27}) yield
$c_{5}=0$. The dispersion relation, originated from the determinant
of the Fourier analyzed form, gives the real part
\begin{equation}{\setcounter{equation}{1}}\label{38}
A_1(z)k^4+A_2(z,\omega)k^3+A_3(z,\omega)k^2+A_4(z,\omega)k+A_5(z,\omega)=0
\end{equation}
which provides four roots, all are imaginary. The imaginary part of
the dispersion relation
\begin{eqnarray}\label{39}
&&B_1(z)k^5+B_2(z,\omega)k^4+B_3(z,\omega)k^3+B_4(z,\omega)k^2+B_5(z,\omega)k\nonumber\\
&&+B_6(z,\omega)=0
\end{eqnarray}
yields five roots of $k$, three real and two complex. The real roots
indicate wave propagation in the region $4\leq z\leq10$ and
$0\leq\omega\leq 10$, shown in Figures \textbf{6}-\textbf{8}. The
waves are directed towards the event horizon given in Figure
\textbf{6}, move away from the event horizon as shown in Figures
\textbf{7} and \textbf{8}. The dispersion is normal as well as
anomalous at random points in Figures \textbf{6} and \textbf{8},
while Figure \textbf{7} indicates normal dispersion in the whole
region. The tables III and IV summarize the results deduced from
these figures.
\begin{figure}
\begin{tabular}{cc}\\
\epsfig{file=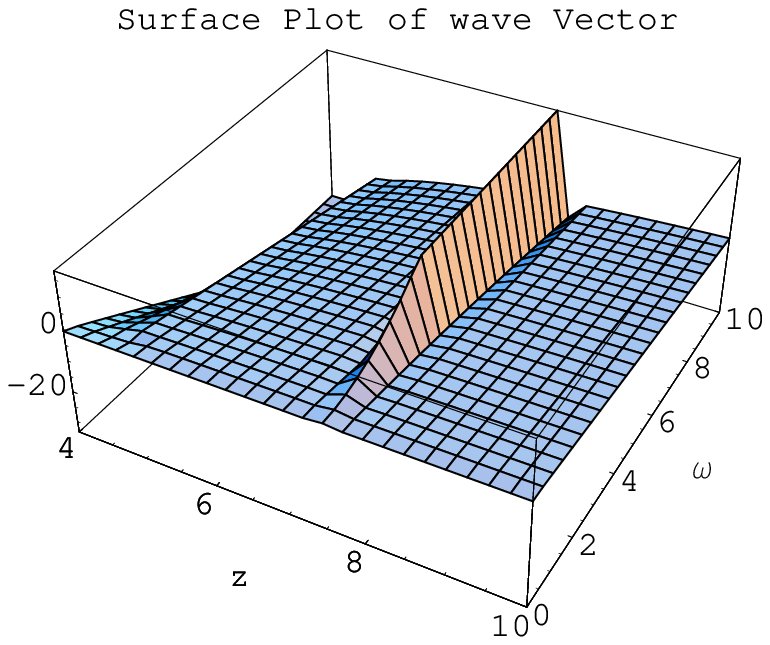,width=0.34\linewidth}
\epsfig{file=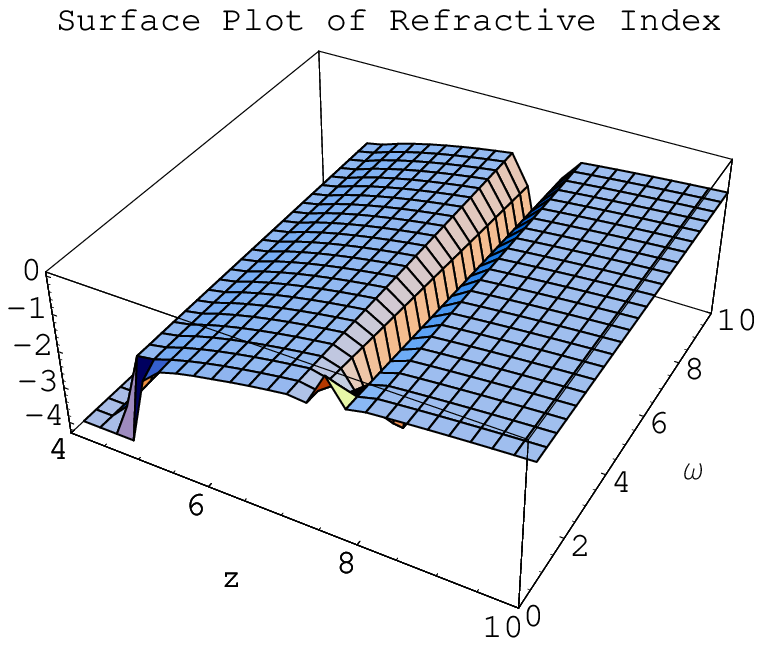,width=0.34\linewidth}\\
\epsfig{file=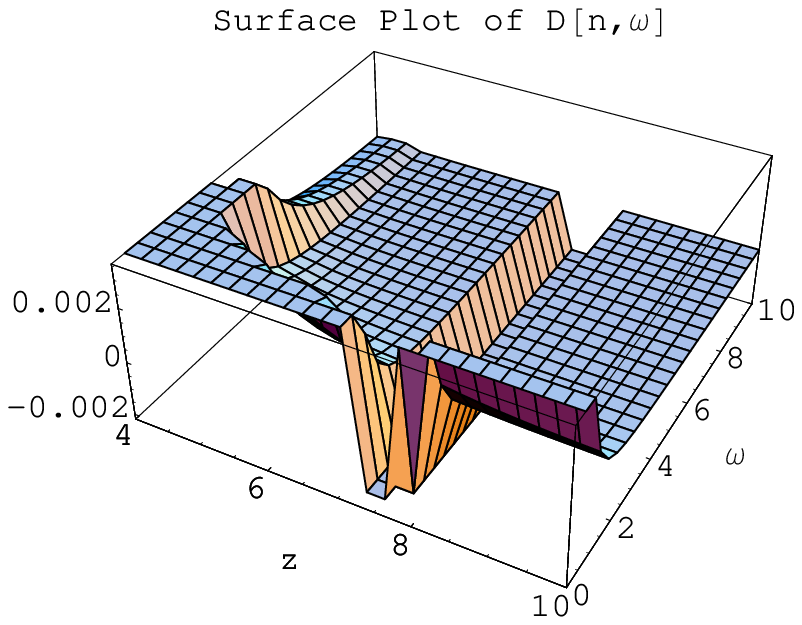,width=0.34\linewidth}
\epsfig{file=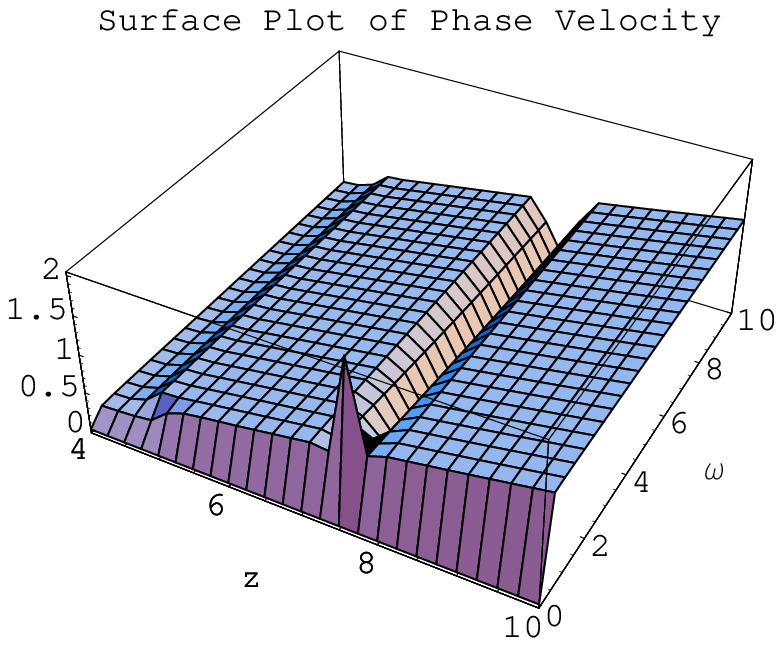,width=0.34\linewidth}
\epsfig{file=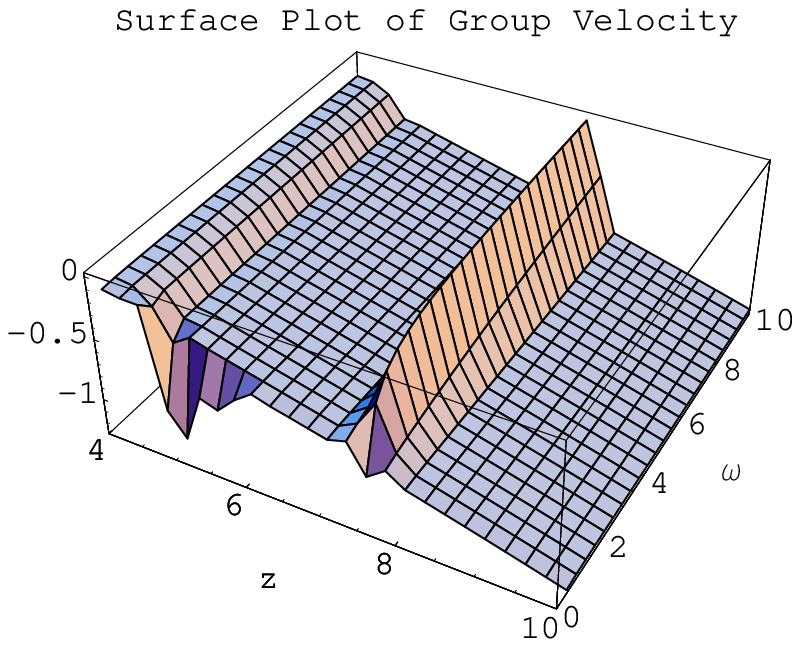,width=0.34\linewidth}\\
\end{tabular}
\caption{Normal and anomalous dispersion of waves is observed.}
\begin{tabular}{cc}
\epsfig{file=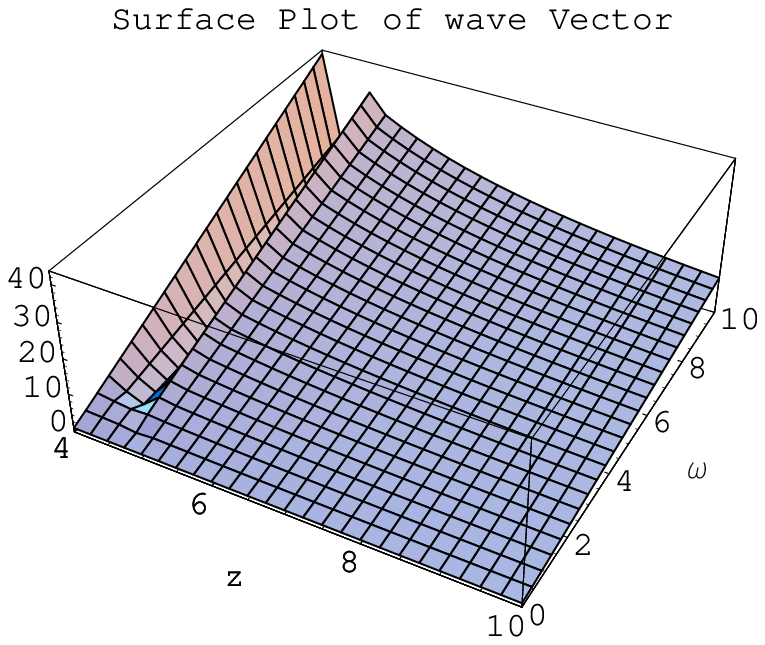,width=0.34\linewidth}
\epsfig{file=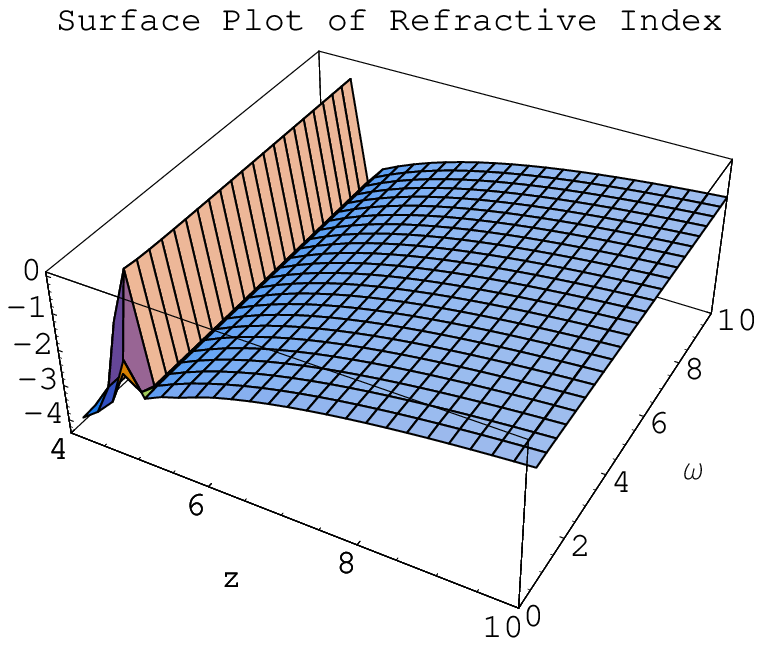,width=0.34\linewidth}\\
\epsfig{file=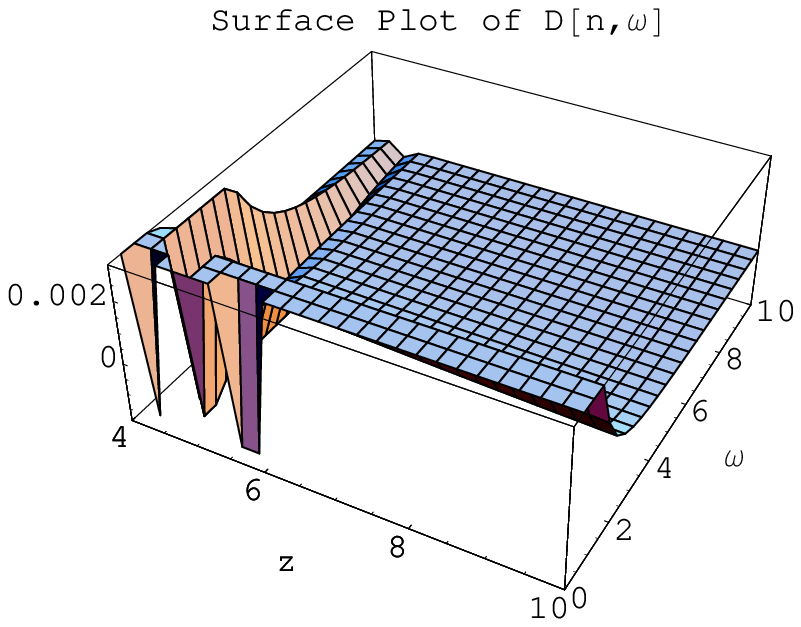,width=0.34\linewidth}
\epsfig{file=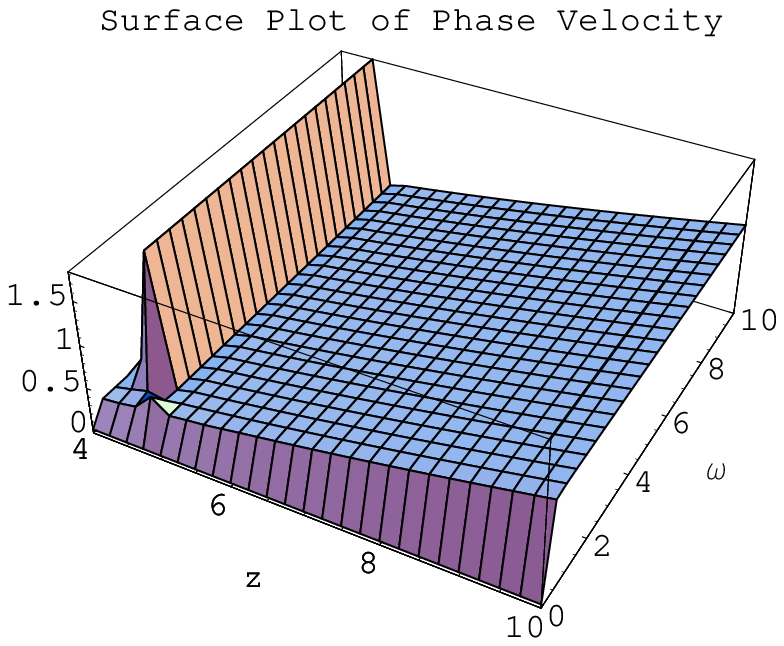,width=0.34\linewidth}
\epsfig{file=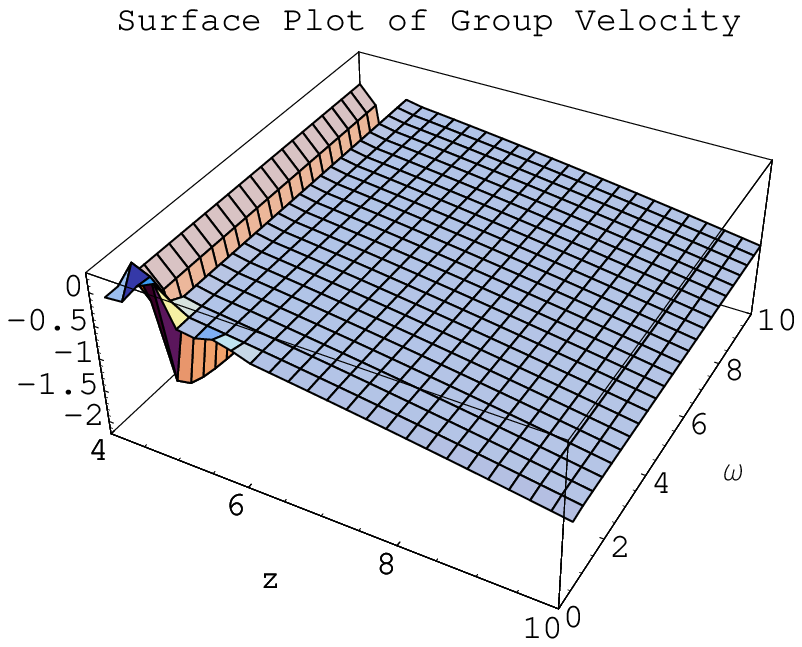,width=0.34\linewidth}\\
\end{tabular}
\caption{Waves disperse normally in the whole region.}
\end{figure}
\begin{center}
Table III. Direction and refractive index of waves
\end{center}
\begin{center}
\begin{tabular}{|c|c|c|c|c|}
\hline\textbf{Fig.} & \textbf{Direction of Waves} &
\textbf{Refractive Index} ($n$)\\ \hline
& & $n<1$ and decreases in the region \\
\textbf{6} & Move towards the event horizon & $6.95\leq z\leq
7.3,0\leq\omega\leq 10$\\&& with the decrease in $z$  \\
\hline
& & $n<1$ and increases in the region\\
\textbf{7} & Move away from the event horizon & $4.3\leq z\leq 4.7,
0\leq\omega\leq
10$\\&&with the decrease in $z$  \\
\hline
& & $n<1$ and increases in the region\\
\textbf{8} & Move away from the event horizon & $4.4\leq z\leq 4.9,
0\leq\omega\leq
10$\\&&with the decrease in $z$  \\
\hline
\end{tabular}
\end{center}
\begin{figure}
\begin{tabular}{cc}\\
\epsfig{file=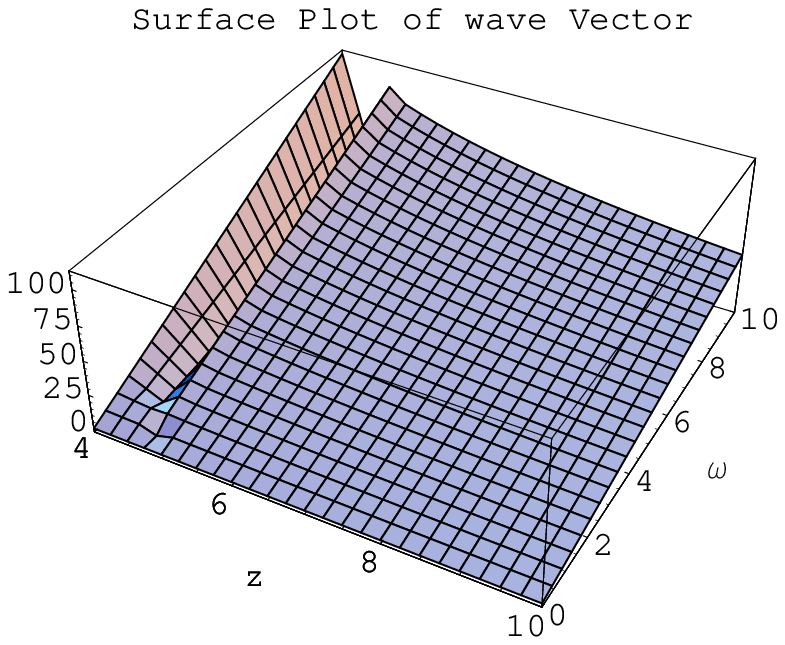,width=0.34\linewidth}
\epsfig{file=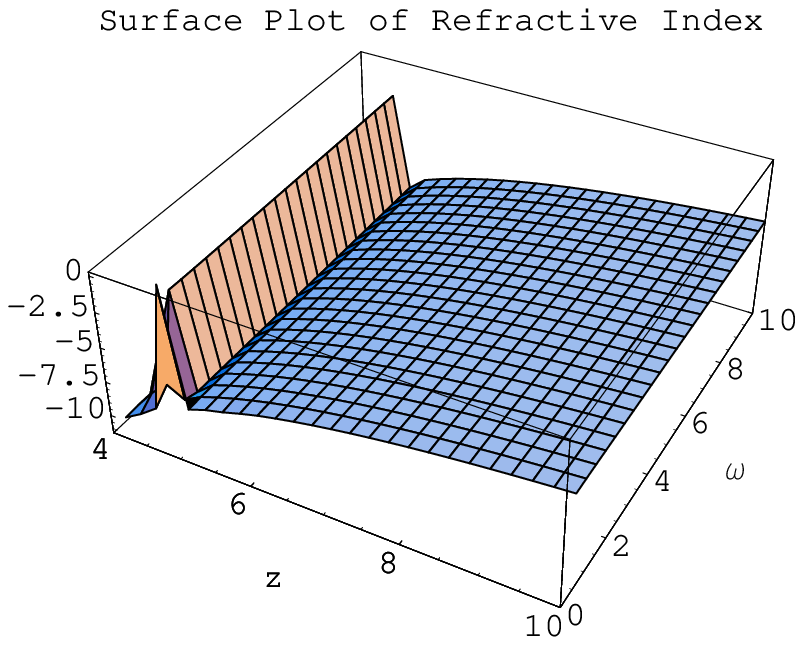,width=0.34\linewidth}\\
\epsfig{file=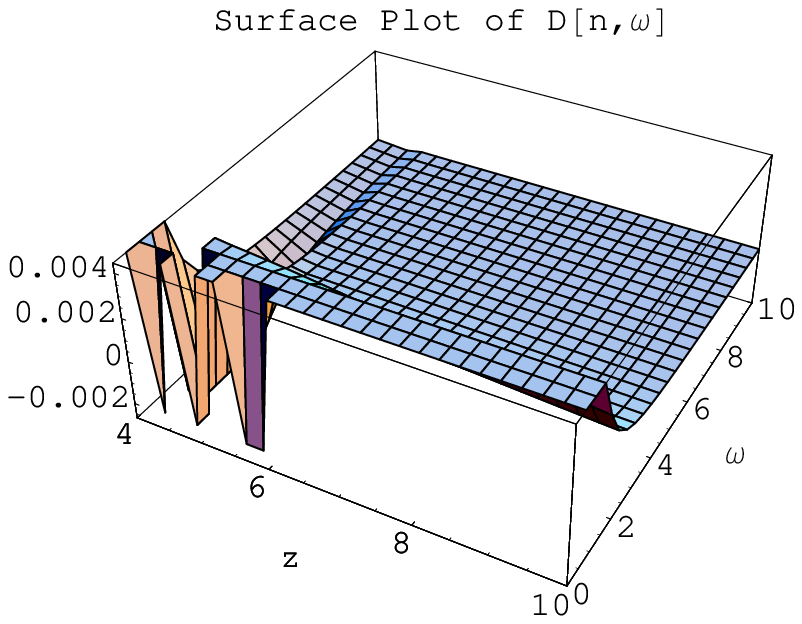,width=0.34\linewidth}
\epsfig{file=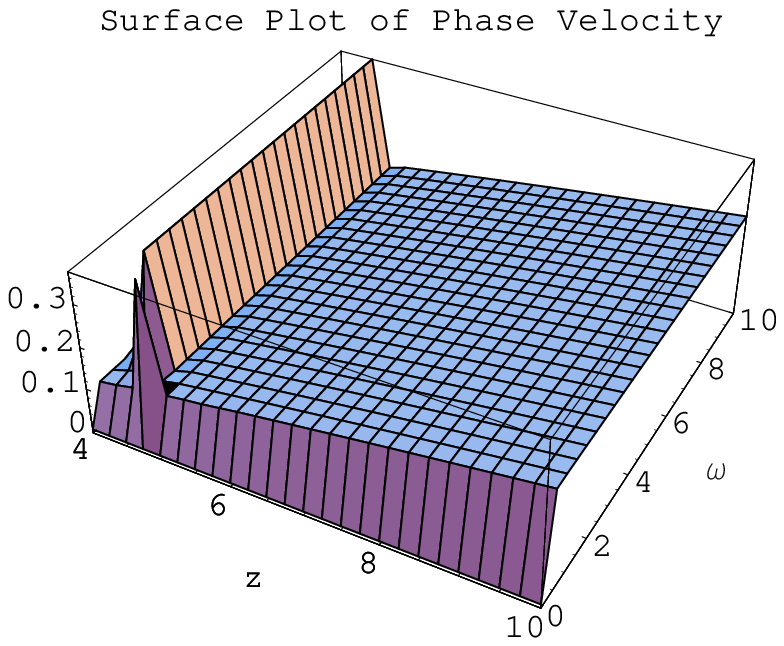,width=0.34\linewidth}
\epsfig{file=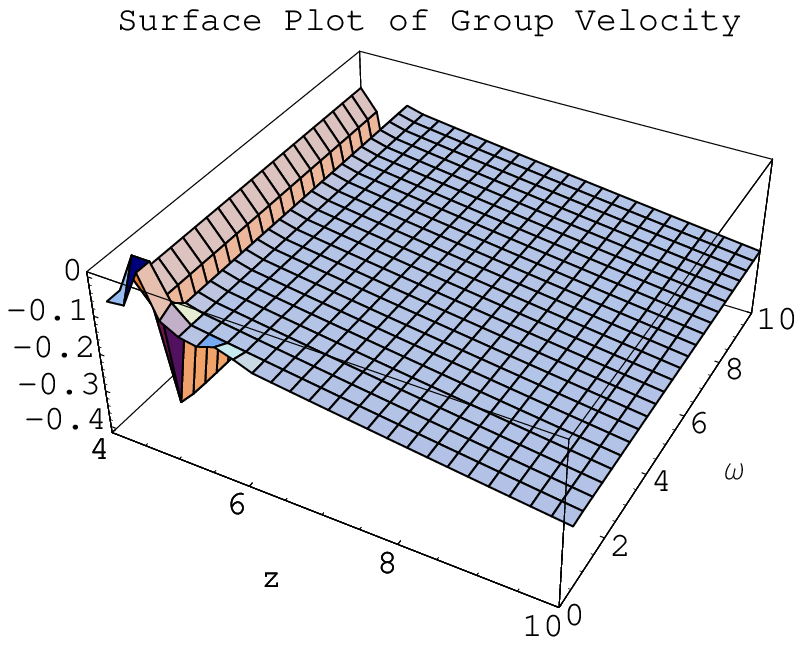,width=0.34\linewidth}\\
\end{tabular}
\caption{Waves exhibit normal as well as anomalous dispersion at random
points.}
\end{figure}

The regions of normal and anomalous dispersion for the roots
exhibiting random dispersion can be separated as follows:

\begin{center}
Table IV. Regions of random dispersion
\end{center}
\begin{center}
\begin{tabular}{|c|c|c|c|c|}
\hline \textbf{Fig.} & \textbf{ Normal dispersion} &
\textbf{Anomalous dispersion}
\\\hline & $4\leq z\leq 7, 2\leq\omega\leq 10$      &
$7\leq z\leq 7.8, 1\leq\omega\leq 4$ \\
\textbf{6} & $7.9\leq z\leq 10, 3\leq\omega\leq 10$   &
\\\hline
& $5.7\leq z\leq 10, 0.5\leq\omega\leq 10$  &
$4\leq z\leq 4.6, 0.03\leq\omega\leq 0.35$     \\
\textbf{8}&   & $4.7\leq z\leq 4.9, 0.45\leq\omega\leq 1.6$
\\\hline
\end{tabular}
\end{center}

\section{Summary}

This paper is devoted to the study of dispersion modes for hot
plasma surrounding SdS black hole in a Veselago medium. For the
formulation of GRMHD equations in this unusual medium, we use the
ADM $3+1$ formalism. The strong gravitational field of black hole
disturbs the flow of surrounding plasma. To determine the impression
of black hole gravity on plasma flow, we have applied linear
perturbations to the GRMHD equations. Further, we have assumed that
plasma flow is in two dimensions and evaluated the component form of
linearly perturbed GRMHD equations. Finally, we have established
dispersion relations for the rotating (non-magnetized and
magnetized) plasma with the help of Fourier analysis techniques.

In rotating non-magnetized plasma, waves move away from the event
horizon shown in Figures \textbf{2} and \textbf{5} while waves are
directed towards the event horizon given in Figures \textbf{1},
\textbf{3} and \textbf{4}. The dispersion is found to be normal in
Figures \textbf{2}, \textbf{3} and \textbf{5} in the whole region.
It can be noticed that Figures \textbf{1} and \textbf{4} exhibit
normal as well as anomalous dispersion at random points. For
rotating magnetized background, waves are directed towards the event
horizon in Figure \textbf{6} while Figures \textbf{7} and \textbf{8}
indicate that waves moves away from the event horizon. The
dispersion is randomly distributed (normal as well as anomalous)
shown in Figures \textbf{6} and \textbf{8} while Figure \textbf{7}
admits normal dispersion in the whole region.

In the usual medium, the refractive index is always greater than one
but for Veselago medium its value must be less than one. The graphs
of refractive index in all figures show that region has refractive
index less than one and increases or decreases in small regions.
Also, the phase velocity is greater than group velocity for both
non-magnetized and magnetized backgrounds. These are the prominent
features of the Veselago medium which demonstrate the validity of
this unusual medium for both rotating (non-magnetized and
magnetized) background around SdS event horizon.

We have seen that for hot plasma surrounding the Schwarzschild black
hole, most of the waves have anomalous and normal dispersion at
random points while some waves exhibit anomalous dispersion in the
whole region \cite{36}. Moreover, most of the waves
are directed away from the event horizon of the Schwarzschild black
hole. However, here we have found that dispersion is normal at most
of the points. Thus it can be established that more information can
be extracted from magnetosphere by the inclusion of positive
cosmological constant in non-rotating black hole. Also, the outcome
of SdS black hole for isothermal plasma in a Veselago medium \cite{37}
shows wave propagation in different regions, i.e.,
$-5\leq z\leq-1,~0\leq\omega\leq10$ and $1\leq
z\leq5,~0\leq\omega\leq10$. For the isothermal plasma, some waves
move towards event horizon and other move towards outer end of
magnetosphere. This comparison shows that the wave properties can be
better understood for the highly ionized, i.e., hot plasma around
SdS magnetosphere.
\renewcommand{\theequation}{A\arabic{equation}}

\section*{Appendix A1}

The Maxwell equations for such a medium are
\begin{eqnarray}{\setcounter{equation}{1}}
\label{40}&&\nabla.\textbf{B}=0,\\
\label{41}&&\nabla\times\textbf{E}+\frac{\partial\textbf{B}}{\partial
t}=0,\\\label{42}
&&\nabla\cdot\textbf{E}=-\frac{\rho_e}{\epsilon},\\
\label{43}&&\nabla\times\textbf{B}=-\mu\textbf{j}+\frac{\partial\textbf{E}}{\partial
t}=0.
\end{eqnarray}
The GRMHD equations for the SdS spacetime in Rindler coordinates
become
\begin{eqnarray}\label{49}
&&\frac{\partial\textbf{B}}{\partial t}=-\nabla \times(\alpha
\textbf{V}\times\textbf{B}),\\\label{50}
&&\nabla.\textbf{B}=0,\\\label{51} &&\frac{\partial\rho_0}{\partial
t}+(\alpha\textbf{V}.\nabla)\rho_0+\rho_0\gamma^2
\textbf{V}.\frac{\partial\textbf{V}}{\partial
t}+\rho_0\gamma^2\textbf{V}.(\alpha\textbf{V}.\nabla)\textbf{V}\nonumber\\
&&+\rho_0{\nabla.(\alpha\textbf{V})}=0, \\\label{52}
&&\{(\rho_0\mu\gamma^2+\frac{\textbf{B}^2}{4\pi})\delta_{ij}
+\rho_0\mu\gamma^4V_iV_j
-\frac{1}{4\pi}B_iB_j\}(\frac{1}{\alpha}\frac{\partial}{\partial
t}+\textbf{V}.\nabla)V^j\nonumber\\
&&-(\frac{\textbf{B}^2}{4\pi}\delta_{ij}-\frac{1}{4\pi}B_iB_j)
V^j,_kV^k+\rho_0\gamma^2V_i\{\frac{1}{\alpha}\frac{\partial
\mu}{\partial t}+(\textbf{V}.\nabla)\mu\}\nonumber\\
&&=-\rho_0\mu\gamma^2a_i-p,_i+
\frac{1}{4\pi}(\textbf{V}\times\textbf{B})_i\nabla.(\textbf{V}\times\textbf{B})
-\frac{1}{8\pi\alpha^2}(\alpha\textbf{B})^2,_i\nonumber\\
&&+\frac{1}{4\pi\alpha}(\alpha B_i),_jB^j-\frac{1}{4\pi\alpha}
[\textbf{B}\times\{\textbf{V}\times(\nabla\times(\alpha\textbf{V}
\times\textbf{B}))\}]_i,\\
\label{53}&&(\frac{1}{\alpha}\frac{\partial}{\partial
t}+\textbf{V}.\nabla)(\mu\rho_0\gamma^2)-\frac{1}{\alpha}\frac{\partial
p }{\partial
t}+2\mu\rho_0\gamma^2(\textbf{V}.\textbf{a})+\mu\rho_0\gamma^2
(\nabla.\textbf{V})\nonumber\\&&-\frac{1}{4\pi}
(\textbf{V}\times\textbf{B}).(\textbf{V}\times\frac{1}{\alpha}\frac{\partial
\textbf{B}}{\partial t})-\frac{1}{4\pi}
(\textbf{V}\times\textbf{B}).(\textbf{B}\times\frac{1}{\alpha}\frac{\partial
\textbf{V}}{\partial
t})\nonumber\\&&+\frac{1}{4\pi\alpha}(\textbf{V}\times\textbf{B}).
(\nabla\times\alpha\textbf{B})=0.
\end{eqnarray}

\end{document}